\documentclass[article,twocolumn,english,footinbib,tightenlines,nobibnotes,aps,prb,superscriptaddress]{revtex4-1}
\usepackage[usenames,dvipsnames]{color}
\usepackage{amsmath}
\usepackage{amsfonts}
\usepackage{amssymb}
\usepackage{mathtools}
\usepackage{grffile}
\usepackage[caption=false]{subfig}
\usepackage{graphicx,xcolor}
\usepackage{mathtools}
\usepackage{paralist}
\usepackage{grffile}
\usepackage{pict2e}
\usepackage{mathtools}
\usepackage{hyperref}
\usepackage{bbding}
\usepackage{epstopdf}
\usepackage{bezier} 
\usepackage{bm}
\usepackage{subfloat}
\usepackage{balance}
\usepackage{soul}
\usepackage{multirow}
\usepackage{babel}

\usepackage{times}
\newcommand{\mib}{\bm}

\def\CL{{\mathcal L}}
\def\RPsi{{\rm\Psi}}
\def\ns{^{\vphantom{*}}}
\def\uar{\uparrow}
\def\dar{\downarrow}
\def\sket#1{{|   #1   \rangle}}
\def\sbra#1{{\langle \,  #1 \, |}}
\def\ket#1{{\big| \, #1\, \big\rangle}}

\def\frac#1#2{{\textstyle{#1 \over #2}}}
\def\half{\frac{1}{2}}
\def\third{\frac{1}{3}}
\def\yd{^\dagger}
\def\nd{^{\vphantom{\dagger}}}
\def\ns{^{\vphantom{*}}}
\def\Tra{\mathop{{\rm Tr}}\,}

\def\eps{\epsilon}
\def\ve{\varepsilon}
\def\sss#1{{\scriptscriptstyle #1}}

\def\ssr#1{{\sss{\rm #1}}}

\def\Ba{{\mib a}}
\def\Bb{{\mib b}}
\def\Bk{{\mib k}}
\def\BR{{\mib R}}
\def\BS{{\mib S}}
\def\BK{{\mib K}}

\def\BM{{\mib M}}
\def\Bq{{\mib q}}
\def\xhat{\hat{\mib x}}
\def\yhat{\hat{\mib y}}
\def\zhat{\hat{\mib z}}
\def\nhat{\hat{\mib n}}
\def\ehat{\hat{\mib e}}
\def\Rc{{\rm c}}

\def\RS{{\rm S}}
\def\Bsigma{{\mib \sigma}}
\def\ie{{\it i.e.\/}}
\mathchardef\Gamma="7100

\begin{document}

\title{Order and disorder in ${\rm SU}(N)$ simplex solid antiferromagnets}
\author{Yury Yu. Kiselev}
\affiliation{Department of Physics, University of California at San Diego, La Jolla CA 92093, USA}
\author{S. A. Parameswaran}
\affiliation{Department of Physics and Astronomy, University of California, Irvine CA 92697, USA}
\author{Daniel P. Arovas}
\affiliation{Department of Physics, University of California at San Diego, La Jolla CA 92093, USA}
\date{\today}
\begin{abstract}

We study the structure of quantum ground states of simplex solid models, which are generalizations of the valence bond construction
for quantum antiferromagnets originally proposed by Affleck, Kennedy, Lieb, and Tasaki (AKLT) [{\sl Phys. Rev. Lett.} {\bf 59}, 799
(1987)].  Whereas the AKLT states are created by application of bond singlet operators for ${\rm SU}(2)$ spins, the simplex solid construction
is based on $N$-simplex singlet operators for ${\rm SU}(N)$ spins.  In both cases, a discrete one-parameter family of translationally-invariant
models with exactly solvable ground states is defined on any regular lattice, and the equal time ground state correlations are given by
the finite temperature correlations of an associated classical model on the same lattice, owing to the product form of the wave functions
when expressed in a ${\rm CP}^{N-1}$ coherent state representation. We study these classical companion models via a mix of Monte Carlo simulations, mean-field arguments, and low-temperature effective field theories. Our analysis reveals that the ground states of ${\rm SU}(4)$ edge- and ${\rm SU}(8)$ face-sharing cubic lattice simplex solid models  
are long range ordered for sufficiently large values of the discrete parameter, whereas the ground states of the ${\rm SU}(3)$ models on the kagome (2D) and hyperkagome (3D) lattices are always quantum disordered. The kagome simplex solid exhibits strong local order absent in its three-dimensional hyperkagome counterpart, a contrast that we rationalize with arguments similar to those leading to `order by disorder'.

\end{abstract}
\maketitle
\section{Introduction}
The study of quantum magnetism is an enduring theme in condensed matter physics, particularly in the search for new phases of matter. Much of our understanding of classical and quantum order and associated critical phenomena has been bolstered by studies of magnetism in diverse situations. In this context, a special role is played by so-called {\it quantum paramagnets}: gapped zero-temperature ground states that retain all symmetries of the high-temperature phase.  Much recent attention has been devoted to the physics of topological quantum spin liquids, which also exhibit
featureless ground states breaking no symmetries.  However, such phases are marked by a confluence of exotica, including ground state degeneracy on multiply connected spaces,
elementary excitations with fractional statistics, and nonlocal quantum entanglement.  Quantum paramagnets, by contrast, have nondegenerate ground states, bosonic bulk elementary
excitations, and their entanglement entropy obeys a conventional area law, with no universal subleading topological contributions.

In a landmark paper\cite{PhysRevLett.59.799} , Affleck, Kennedy, Lieb and Tasaki (AKLT) presented an explicit construction of a family of `valence bond solid' (VBS) wavefunctions,
each specified by a lattice $\CL$ (which we will assume to be regular, \ie\ all sites have the same coordination number), and a positive integer $M$.  The local spin quantum number $S$
is related to the lattice coordination number $r$
according to $S=\half M r$ (hence large $M$ means less quantum fluctuations) and each VBS wave function is a ground state of a Hamiltonian which may be written as a sum of
local projection operators.  The simplest example is the one-dimensional $S=1$ AKLT chain, whose wavefunction is the nondegenerate ground state (on a ring) of the Hamiltonian
$H=\sum_n \BS_n\cdot\BS_{n+1} + \frac{1}{3}(\BS_n\cdot\BS_{n+1})^2$.  This state provided the first exact wavefunction for a system exhibiting a Haldane gap\cite{Haldane1983464,FDMH1988}. 
These isotropic valence bond solid (VBS) states provide a useful paradigm for  quantum paramagnets in which both spin and lattice point group symmetries remain unbroken. 
As noted more recently by Yao and Kivelson~\cite{YaoKivelson}, the AKLT states are also examples of `fragile Mott insulators', that cannot be adiabatically connected to a band insulator
while preserving certain point-group symmetries.  In dimensions $d>2$, the VBS states may exhibit long-ranged N{\'e}el order if $M$ is sufficiently large.

The AKLT construction is based on application of local singlet bond operators, and may be visualized as a Tinkertoy network.  In this paper, we shall explore the properties of an extension
of the VBS family to SU($N$) quantum spins, first discussed by one of us\cite{arovas:104404}, in which the singlets reside on $N$-site simplices.  These `simplex solid' models have much
in common with their VBS relatives, including featureless $T=0$ quantum paramagnetic phases, parent Hamiltonians which are sums of local projectors, nondegenerate ground
states regardless of base space topology, area law entanglement, and possibly broken SU($N$) symmetry in $d>2$ dimensions.  As with the VBS states, for each lattice $\CL$
there is a discrete one-parameter family of models, labeled by an integer $M$, which together determine the local representation of SU($N$).  And, similarly, in dimensions $d>2$ the
simplex solids may solidify into a spin crystal, {\it i.e.\/} a generalized N{\'e}el state, provided $M$ is sufficiently large.

Here, we consider several examples of simplex solid states, and study
their properties via a combination of methods, including (classical!) Monte Carlo and various analytical methods.
The key technical feature which permits such analyses is a mapping, via generalized spin coherent states, of the equal time quantum correlations of the simplex solid wavefunctions to
finite temperature correlations of an associated {\it classical\/} model on the same lattice -- another aspect shared with the VBS states\cite{PhysRevLett.60.531}.  We will therefore focus on the
properties of these classical models, their possible ordered phases, mean field descriptions, and analysis of low-energy effective models.
We will find, and explain why, that, unlike the VBS models, some $d>2$ simplex solids never order for {\it any\/} finite value of $M$, no matter how large.  Another noteworthy feature is that
local and long-ranged order in simplex solids is due to an order-by-disorder mechanism\footnote{OBD1980}.
The simplex solid states furnish a new paradigm for SU($N$) quantum magnetism, and have recently been extended to more general tensor network constructions\cite{PhysRevX.4.011025},
which may provide useful compact ways to express trial state wavefunctions for interacting quantum systems.


\section{Valence bond and Simplex solid states}
As mentioned above, for each lattice $\CL$ there is a family of VBS states indexed by a positive integer $M$, constructed as follows\cite{PhysRevLett.59.799}: First,
place $Mr$ spin-$\half$ objects on each site, where $r$ is the lattice coordination number (we assume $r\ns_i=r$ for all sites $i$ in $\CL$).  Next, contract the SU(2) indices by forming $M$
singlet bonds on each link of the lattice.  Finally, symmetrize over all the SU(2) indices on each site.  This last step projects
each site spin into the totally symmetric $S=\half M r$ representation, \ie\ a Young tableaux with one row of $Mr$ boxes.
The general state $\sket{\RPsi(\CL,M)}$ is conveniently
represented using the Schwinger boson construction \cite{PhysRevLett.60.531}, where $\BS=b\yd_\mu\,\Bsigma\ns_{\mu\nu}\,b\nd_\nu$
and the total boson number on each site is $b\yd_\uar b\ns_\uar + b\yd_\dar b\nd_\dar=2S$\,:
\begin{equation}
\ket{\RPsi(\CL,M)}=\prod_{\langle ij\rangle\in\CL}
\big(b\yd_{i\uar}\,b\yd_{j\dar} - b\yd_{i\dar}\,b\yd_{j\uar}\big)^M\,\ket{0}\ .
\label{VBS}
\end{equation}
Since the bond operator $\phi\yd_{ij}=\ve^{\mu\nu}\,b\yd_{i\mu}\,b\yd_{j\nu}$ transforms as an SU(2) singlet, $M$ of the bosons at site $i$
are fully entangled in a singlet state with $M$ bosons on site $j$, so that
the maximum value of the total spin $J\ns_{ij}$ is $2S-M$, and thus 
 $\sket{\RPsi(\CL,M)}$ is an exact zero energy ground state for any
Hamiltonian of the form $H=\sum_{\langle ij\rangle} \sum_{J=2S-M+1}^{2S} V\ns_J\,P\ns_J(ij)$, where $V\ns_J>0$ are pseudopotentials and
$P\ns_J(ij)$ is the projector onto total spin $J$ for the link $(ij)$.

Many properties of $\sket{\RPsi(\CL,M)}$ may be gleaned from its coherent state
representation~\cite{PhysRevLett.60.531},  $\RPsi\ns_{\!\CL,M}[z]=\prod_{\langle ij\rangle\in\CL} \big(\eps^{\mu\nu}\,z\ns_{i\mu}\,z\ns_{j\nu}\big)^{\!M}$, where for each
site $i$, $z\ns_i$ is a rank-2 spinor with $z\yd_i z\nd_i=1$ and $z\ns_i\equiv e^{i\alpha}z\ns_i$, \ie\ an element of the complex
projective space ${\rm CP}^1\cong \RS^2$.  In particular, one has $\big|\RPsi\ns_{\!\CL,M}[z]\big|^2=e^{-H\ns_{\rm cl}/T}$, where
\begin{equation}
H\ns_{\rm cl}=-\!\!\sum_{\langle ij\rangle \in \CL} \!\ln\!\bigg({1-\nhat\ns_i\cdot\nhat\ns_j\over 2}\bigg)\ ,
\end{equation}
with $\nhat\ns_i=z\yd_i\,\Bsigma\,z\nd_i\in\RS^2$ is a unit vector, is the Hamiltonian for a classical ${\rm O}(3)$ antiferromagnet
on the same lattice $\CL$, and $T=1/M$ is a fictitious temperature. This is analogous to Laughlin's `plasma analogy' for the fractional quantum Hall effect, and  we may similarly use well-known results in classical statistical mechanics to deduce properties of the state described by $\sket{\RPsi(\CL,M)}$. Specifically, we may invoke the Hohenberg-Mermin-Wagner theorem to conclude
that all 
AKLT states in dimensions $d\le 2$ lack long-range magnetic order 
 since they correspond to a classical ${\rm O}(3)$ system at finite
temperature on the same lattice\footnote{Note however that in two dimensions there remains the possibility of broken point-group symmetry without spin order -- as these are discrete symmetries, their breaking is {\it not} forbidden in $d=2$.}.  For $d>2$, a mean-field analysis \cite{PhysRevLett.59.799,parameswaran:024408} suggests that the AKLT
states on bipartite lattices possess long-ranged two sublattice antiferromagnetic order for $T<T_\Rc^{\ssr{MF}}=\third r$, \ie\ 
$M>M_\Rc^{\ssr{MF}}=3 r^{-1}$.  Since the minimum possible value for $M$ is $M=1$, the mean field analysis suggests that all such $d=3$
models, where $r>3$, are Neel ordered. However, mean field theory famously fails to account for fluctuation effects which drive $T\ns_\Rc$
lower -- hence $M\ns_\Rc$ higher --
for instance, ref.~\onlinecite{parameswaran:024408} found, using classical Monte Carlo simulations  of 
the corresponding classical ${\rm O}(3)$ model, that the $S=2$ (\ie\ $M=1$)
AKLT  state on the diamond lattice ($r=4$) is quantum-disordered.  Ref.~\onlinecite{parameswaran:024408} also showed that AKLT states on the frustrated pyrochlore lattice were quantum-disordered for $S\leq 15$ (at least). A subsequent extension of the AKLT model to locally tree-like graphs  -- often used to model disordered systems -- found AKLT states that exhibit not only long-range order and quantum disorder, but also those that showed spin glass-like order for large values of the singlet parameter and/or the local tree coordination number\cite{Laumann:2010p1}.

Upon enlarging the symmetry group of each spin to ${\rm SU}(N)$, there are two commonly invoked routes to singlet ground states. The first is to work exclusively with bipartite
lattices, and choose the 
spins on  
one sublattice to transform according to the ($N$-dimensional) fundamental representation of ${\rm SU}(N)$,
while those on the 
other transform according to the ($N$-dimensional) conjugate representation.  One then has
$N\otimes{\bar N}=\bullet\oplus{\rm adj}$\,, where $\bullet$ denotes the singlet and ${\rm adj}$ the $(N^2-1)$-dimensional adjoint
representation.  Proceeding thusly, one can develop a systematic large-$N$ expansion \cite{PhysRevB.38.316,PhysRevLett.62.1694}. Note, however, that on bipartite lattices in which the two sublattices are equivalent, most assignments of bond singlets explicitly break either translational or point-group symmetries. (The exceptions typically involve fractionalization, and hence also do not satisfy our desiderata for a featureless quantum paramagnet.)

The second approach, and our exclusive focus in the remainder, is to retain the {\it same} representation of ${\rm SU}(N)$ on each site, but to create singlets which extend over a group of
$N$ sites. (Readers may recognize a family resemblance with the three-quark ${\rm SU}(3)$ color singlet familiar from quantum chromodynamics.)  In this paper, we shall explore the ordered and disordered phases in a class of wave functions which generalize the AKLT valence bond
construction from ${\rm SU}(2)$ to ${\rm SU}(N)$, and from singlets on bonds to those over simplices.  The construction and analysis of these ``simplex solids" \cite{arovas:104404} parallels what we know about the
AKLT states.
 If $\Gamma$ denotes an $N$ site simplex 
(henceforth an $N$-simplex) whose sites are labeled $\{i\ns_1,\ldots,i\ns_N\}$, then the operator
\begin{equation}
\phi\yd_\Gamma=\ve^{\alpha\ns_1\cdots\alpha\ns_N}\,b\yd_{i\ns_1\alpha\ns_1}\cdots b\yd_{i\ns_N\alpha\ns_N}
\end{equation}
where $b\yd_{i\alpha}$ creates a Schwinger boson of flavor $\alpha$ on site $i$, transforms as an ${\rm SU}(N)$ singlet. Generalizing the
product over links in the AKLT construction to a product over $N$-simplices, one arrives at the simplex solid state \cite{arovas:104404},
\begin{equation}\label{eq:ssdef}
\ket{\RPsi(\CL,M)}=\prod_{\Gamma\in\CL} \big(\phi\yd_\Gamma\big)^{\!M}\ket{0}\ .
\end{equation}
The resulting local representation of ${\rm SU}(N)$ is the symmetric one described by a Young table with one row and $p=M\zeta$ boxes, where $\zeta$ 
is the number of simplices to which each site on $\CL$ belongs, a generalization of the lattice coordination number $r$ in the case $N=2$.
Projection operator Hamiltonians which render the simplex solid (SS) states exact zero energy ground states were discussed in Ref.~\onlinecite{arovas:104404}. Written in terms of the $N$-flavors of Schwinger bosons, the ${\rm SU}(N)$ spin operators take the form
$S\ns_{\alpha\beta}=b\yd_\alpha\,b\nd_\beta-{p\over N}\delta\ns_{\alpha\beta}$, and satisfy the commutation relations
$\big[S\ns_{\alpha\beta}\,,\,S\ns_{\mu\nu}\big]=\delta\ns_{\beta\mu}\,S\ns_{\alpha\nu}- \delta\ns_{\alpha\nu}\,S\ns_{\beta\mu}.$ 
As in the AKLT case, while the wave functions ~\eqref{eq:ssdef} are certainly exact ground states of local parent Hamiltonians, it is imperative to verify that they do in fact describe featureless paramagnets. In addressing this question, it is once again convenient to employ a coherent-state representation (suitably generalized to ${\rm SU}(N)$) so that the answer can be inferred  from analysis of a finite-temperature classical statistical mechanics problem. Using this mapping, described in detail below, in conjunction with the Hohenberg-Mermin-Wagner theorem, we find that although wave functions of the form ~\eqref{eq:ssdef}
preserve all symmetries in one dimension, once again we must entertain the possibility that they exhibit lattice symmetry-breaking but not magnetic order in $d=2$, and that both lattice and spin symmetries are spontaneously broken in $d=3$. 

In $d=2$, we consider the ${\rm SU}(3)$ simplex solid on the kagome lattice, and using a saddle-point free energy estimate and Monte Carlo simulations of the classical model, we show that it remains quantum-disordered for all $M$, although there is substantial local sublattice order, corresponding to the so-called $\sqrt{3}\times\sqrt{3}$ structure, for large $M$ (low effective temperature). We then turn to $d=3$, where we first consider the ${\rm SU}(3)$ simplex solid on the hyperkagome lattice of corner-sharing triangles. Here we find no discernible structure for any $M$, leading us to conclude that {\it all} these simplex solid states are quantum-disordered. We also consider two different simplex solids on the cubic lattice: the ${\rm SU}(4)$ model with singlets on square plaquettes (that share edges), and the  ${\rm SU}(8)$ version with singlets over cubes (that share faces). While the former exhibits long-range order for all $M$ (in other words, the classical companion model has a continuous transition at $T_c>1$),
the latter exhibits long-range order only for $M\geq3$, so that the $M=1,2$ cases are quantum-disordered.

Before proceeding, we briefly comment on related work. Other generalized Heisenberg models have been discussed in
 a variety of contexts. Affleck {\it et al.\/} \cite{Affleck1991467} investigated extended valence bond solid models with exact ground states which break
charge conjugation (${\cal C}$) and lattice translation ($t$) symmetries, but preserve the product ${\cal C}t$.  Their construction
utilized SU($2N$) spins on each lattice site, with $N=Mr$ an integer multiple of the lattice coordination number $r$, with singlet
operators extending over $r+1$ sites.  Greiter and Rachel \cite{greiter:184441} constructed ${\rm SU}(N)$ VBS chains in the fundamental and
other representations.  Shen \cite{PhysRevB.64.132411} and Nussinov and Ortiz \cite{Nussinov2009977} developed models with resonating
Kekul\'{e} ground states described by products of local ${\rm SU}(N)$ singlets.  Plaquette ground states on two-leg ladders were also discussed
by Chen {\it et al.\/} \cite{PhysRevB.72.214428}.   VBS states are perhaps the simplest example of matrix product and tensor network
constructions \cite{PhysRevLett.75.3537,PhysRevLett.93.040502,PhysRevA.70.060302,2006quant8197P}, and recently the projected
entangled pair state (PEPS) construction was extended by Xie {\it et al.\/} to one involving projected entangled simplices \cite{2013arXiv1307.5696X}. We also note that a different generalization to the group Sp($N$) permits the development of a large-$N$ expansion for doped and frustrated lattices~\cite{PhysRevLett.66.1773}. Perhaps more relevant to our discussion here, Corboz {\it et al.\/}~\cite{PhysRevB.86.041106} studied ${\rm SU}(3)$ and ${\rm SU}(4)$ Heisenberg models on the kagome and checkerboard lattices using the infinite-system generalization of PEPS (iPEPS), concluding that the Hamiltonian {\it at the Heisenberg point} exhibits $\Bq=0$ point-group symmetry-breaking \footnote{Specifically, they argue that the singlets will be formed on one of two inequivalent choices of simplices, such as only on up triangles in the kagome. This is the ${\rm SU}(N)$ analog of the Majumdar-Ghosh state.}. Although their work left open the question of its adiabatic continuity to the exactly solvable point of Ref.~\onlinecite{arovas:104404}, this follows immediately, as the order they discuss is inescapable for a simplex solid where the on-site  spins are (as in their work) in the fundamental representation of ${\rm SU}(N)$.

In addition, there are several other examples of featureless quantum paramagnets in the literature, with more general symmetry groups. Besides the aforementioned work by Yao and Kivelson, fragile Mott insulating phases have been recently examined as possible ground states of aromatic molecules in organic chemistry~\cite{2014arXiv1409.6732M}. Quantum paramagnetic analogs of the fragile Mott insulator for {\it bosonic} systems endowed with a ${\rm U}(1)$ symmetry have also been explored, including those with
very similar `plasma mappings' to classical companion models~\cite{KagomeWannier,HoneycombVoronoi}.
Finally, recent work (involving two of the present authors) has identified situations when featureless quantum paramagnets are incompatible with crystalline symmetries and $U(1)$ charge conservation~\cite{Parameswaran:2013ty}.

\section{Classical model and mean field theory}
We first briefly review some results of Ref.~\onlinecite{arovas:104404}.
Using the ${\rm SU}(N)$ coherent states $\sket{z}={1\over\sqrt{p!}} \big(z\ns_\alpha b\yd_\alpha\big)^p\sket{0}$,
we may again, as with the VBS states, express equal time ground state correlations in the simplex solids in terms of
thermal correlations of an associated classical model on the same lattice.  One finds
$\big|\RPsi\ns_{\!\CL,M}[z]\big|^2=e^{-H\ns_{\rm cl}/T}$, with
\begin{equation}
H\ns_{\rm cl}=-\sum_\Gamma \ln|R\ns_\Gamma|^2\quad,
\end{equation}
where
\begin{equation}
R\ns_\Gamma=\eps^{\alpha\ns_1\cdots\alpha\ns_N}\,z\ns_{i\ns_1\alpha\ns_1}\!\!\cdots z\ns_{i\ns_N\alpha\ns_N}\quad,
\label{RGamma}
\end{equation}
where $\{i\ns_1,\ldots,i\ns_N\}$ label the $N$ sites of the simplex $\Gamma$. The temperature is again $T=1/M$.
Note that the quantity $|R\ns_\Gamma|$ has the interpretation of a volume spanned by the ${\rm CP}^{N-1}$ vectors
sitting on the vertices of $\Gamma$.

To derive a mean field theory, assume that $\CL$ is $N$-partite and is partitioned into $N$ sublattices.  (The partitioning may not be the same in all structural unit
cells, as the distinction between Figs. \ref{kagomeQ0} and \ref{kagomeSqrt3} demonstrates in the case $N=3$.)  For each
site $i$ let  $\sigma(i)\in \{1,\ldots,N\}$ denote the sublattice to which $i$ belongs.  Let $\{\omega\ns_\sigma\}$
denote a set of $N$ mutually orthogonal ${\rm CP}^{N-1}$ vectors. Setting $z\ns_i=\omega\ns_{\sigma(i)}$ defines a fully
ordered state which we will refer to as a Potts state, since it is also a ground state for a (discrete) $N$-state Potts antiferromagnet.
In any Potts state, $|R\ns_\Gamma|=1$ for every simplex $\Gamma$, hence the ground state energy is $E\ns_0=0$.

Next define a real scalar order parameter $m$, akin to the staggered magnetization in an antiferromagnet, such that
\begin{equation}
\langle Q\ns_{\alpha\beta}(i) \rangle = m\,\bigg( P^{\sigma(i)}_{\alpha\beta} -{1\over N}\,\delta\ns_{\alpha\beta}\bigg)\quad,
\end{equation}
where $Q\ns_{\alpha\beta}(i)= z^*_{i\alpha}\,z\ns_{i\beta}- {1\over N}\,\delta\ns_{\alpha\beta}$ is a locally
defined traceless symmetric tensor, and where $P^\sigma=\sket{\,\omega\ns_\sigma}\sbra{\omega\ns_\sigma}$ is the
projector onto $\omega\ns_\sigma$. The system is isotropic when $m=0$, while $m=1$ in the Potts state.  One finds
that the mean field critical value for $M=1/T$ is $M^\ssr{MF}_{\rm c}=(N^2-1)/\zeta$. Note that for $N=2$ and $\zeta=r$
we recover the mean field results for the VBS states, \ie\ $M^\ssr{MF}_{\rm c}=3/r$.  Thus, mean field considerations
lead us to expect more possibilities for quantum disordered simplex solids than for the valence bond solids in dimensions
$d>2$, where almost all the VBS states are expected to have two sublattice Neel order on bipartite lattices. One
remarkable feature of the SS mean field theory is that it apparently {\it underestimates\/} the critical
temperature in models where a phase transition occurs, thus overestimating $M\ns_{\rm c}$. 

Expanding about the fully ordered state, writing
\begin{equation}
z\nd_i=\big(1-\pi\yd_i\pi\nd_i\big)^{1/2}\omega\nd_{\sigma(i)} + \pi\nd_i\quad,
\end{equation}
where $\omega\yd_{\sigma(i)}\pi\nd_i=0$, the low-temperature classical Hamiltonian is
\begin{equation}
H\nd_{\rm LT}=\sum_\Gamma\sum_{i<j}^N \big|\pi\yd_{\Gamma\nd_i}\,\omega\nd_{\sigma(\Gamma\nd_j)} + \omega\yd_{\sigma(\Gamma\nd_i)} 
\pi\nd_{\Gamma\nd_j}\big|^2 + {\cal O}(\pi^3)\quad.
\label{HLT}
\end{equation}
The field $\pi\nd_i$ has $(N-1)$ independent complex components.  If $g(\ve)$ is the classical density of states per site, normalized
such that $\int\limits_0^\infty\!d\ve\>g(\ve)=1$, then
\begin{equation}
\langle\pi\yd_i\pi\nd_i\rangle=(N-1)\, T\!\!\int\limits_0^\infty{d\ve\over\ve}\>g(\ve)\quad.
\end{equation}
Another expression estimating $T\ns_{\rm c}$ is obtained by setting $\langle\pi\yd_i\pi\nd_i\rangle= 1$, beyond which point the fixed
length constraint $z\yd_i z\nd_1=1$ is violated, \ie\ the low temperature fluctuations of the $\pi$ field are too large.
In contrast to the mean field expression for the critical temperature, $T\ns_{\rm c}=\zeta/(N^2-1)$, value of $T\ns_{\rm c}$ as determined
from this criterion depends on the nature of the putative ordered phase, and moreover it vanishes if
$\int\limits_0^\infty d\ve\>\ve^{-1}\, g(\ve)$ diverges.

\subsection{Counting degrees of freedom}\label{cdof}
For our models, which are invariant under global ${\rm U}(N)$ rotations, each site hosts a ${\rm CP}^{N-1}$ vector,
with $2(N-1)$ real degrees of freedom (DOF).  Thus, per $N$-simplex, there are $2N(N-1)$ degrees of freedom.  The group
${\rm U}(N)$ has $N^2$ generators, $N$ of which are diagonal.  These diagonal generators act on the spins by multiplying
each of the $\omega\ns_\sigma$ by a phase, which has no consequence in ${\rm CP}^{N-1}$.  Therefore there are only 
$N(N-1)$ independent generators to account for.  Subtracting this number from the number of DOF per simplex,
we conclude that, in a Potts state, each simplex satisfies $N(N-1)$ constraints.  If our lattice consists of $K$
corner-sharing simplices, then there are $KN(N-1)$ total (real) degrees of freedom: $2N(N-1)$ DOF per simplex times $K$ simplices,
and multiplied by $\half$ since each site is shared by two simplices.  There are an equal number of constraints.
Thus, the na{\"\i}ve Maxwellian dimension of the ground state manifold is $D\ns_\ssr{M}=0$.  However, as we shall see below,
we really have $D\ge 0$, and in some situations, such as for the kagome and hyperkagome models discussed below,
$D>0$.  If the number of zero modes is subextensive, the $T=0$ heat capacity per site should be $C(0)=N-1$ by equipartition.

\section{Monte Carlo Simulations}
We simulate the  classical companion model via Monte Carlo simulations using a  single-spin flip Metropolis algorithm. As mentioned above, our primary interest is in determining the phase diagram of the classical model as a function of the temperature, as this will tell us how the quantum system depends on the discrete parameter
$M =  1/T$ (recall that this determines the on-site representation of ${\rm SU}(N)$ by fixing the number of boxes in the Young diagram in a fully symmetric representation of ${\rm SU}(N)$). The classical degrees of freedom, obtained via the coherent-state mapping, are ${\rm CP}^{N-1}$ spins;  in our simulations, each
${\rm CP}^{N-1}$ spin is represented by an $N$-dimensional complex unit vector $\vec{z}$. The remaining ${\rm U}(1)$ local ambiguity is harmless.

Local updates are made by generating an isotropic $\delta \vec{z}$ whose length is distributed according to a Gaussian. The  local
spin vector is updated to
\begin{equation}
\vec{z}^{\,\prime} = {\vec{z} + \delta \vec{z} \over |\, \vec{z}+ \delta \vec{z}\, |} \quad.
\end{equation}
The standard deviation of the Gaussian distribution is adjusted so that a significant fraction  $(\sim 30 \%)$ of proposed moves are accepted. 

In order to obtain independent samples, we simulated $N_{\text{chain}}$ independent Markov chains, typically of a length of $\sim 10^5-10^6$ Monte Carlo steps per site (MCS). Each chain was initialized  with random initial conditions and evolved until the total energy was well-equilibrated, and the initial portions of the chain before this were discarded. For each chain, we obtained the average values of the various quantities and averaged this across chains to get a single number for each temperature. We estimated the error from the standard deviation of the $N_{\text{chain}}$ independent thread averages. This is free of the usual complications of correlated samples inherent in estimating the error from a single chain, and it frees us of the need to compute autocorrelation times to weight our error estimate. Note that in the lowest-temperature samples, we used a relatively modest number of independent chains $N_{\text{chain}}\lesssim10$, but this was already sufficient to obtain reasonably small error bars.

We analyze two main observables.  The first is the heat capacity $C=\textsf{var}(H_{\rm cl})/T$, proportional to the square of the RMS energy
fluctuations.  The second is a generalized structure factor, which is built from an appropriate tensor order parameter, 
\begin{equation}
Q\ns_{\alpha\beta}(i)=z^*_{i,\alpha}\,z\ns_{i,\beta} - {1 \over N}\, \delta_{\alpha \beta}\quad.
\end{equation}
Note that ${\vec z}\ns_i$ itself cannot be used as an order parameter, because its overall phase is ambiguous.  This ambiguity is eliminated
in the definition of $Q\ns_{\alpha\beta}(i)$, which is similar to the order parameter of a nematic phase.
This tensor has the following properties:
\begin{itemize}
\item $\Tra{Q} = 0$
\item $\langle \,Q\, \rangle\to 0$ as $T\to\infty$ at all sites
\item $\Tra(Q^2) = {N-1 \over N}$
\item $\Tra(QQ') = -{1 \over N}$ if $z\yd z'=0$\quad.
\end{itemize}
Thus, in any Potts state, $\Tra Q(i)\,Q(j)=-{1\over N}$ for any nearest neighbor pair $(ij)$.  A more detailed measure of
order is afforded by the generalized structure factor, which is given by the Hermitian matrix
\begin{equation}
S_{ij}(\Bk) = {1 \over \Omega} \sum_{\BR, \BR'}  e^{i \Bk (\BR - \BR')}\, {\rm Tr}\big[ Q(\BR, i)\, Q(\BR', j) \big]\ ,
\end{equation}
where $\BR$ is a Bravais lattice site, $\Omega$ is the total number of the unit cells, and $i$ and $j$ are sublattice indices.
The rank of $S\ns_{ij}(\Bk)$ is the number of basis vectors in the lattice.

\begin{figure}[t]
\includegraphics[width=0.8\linewidth]{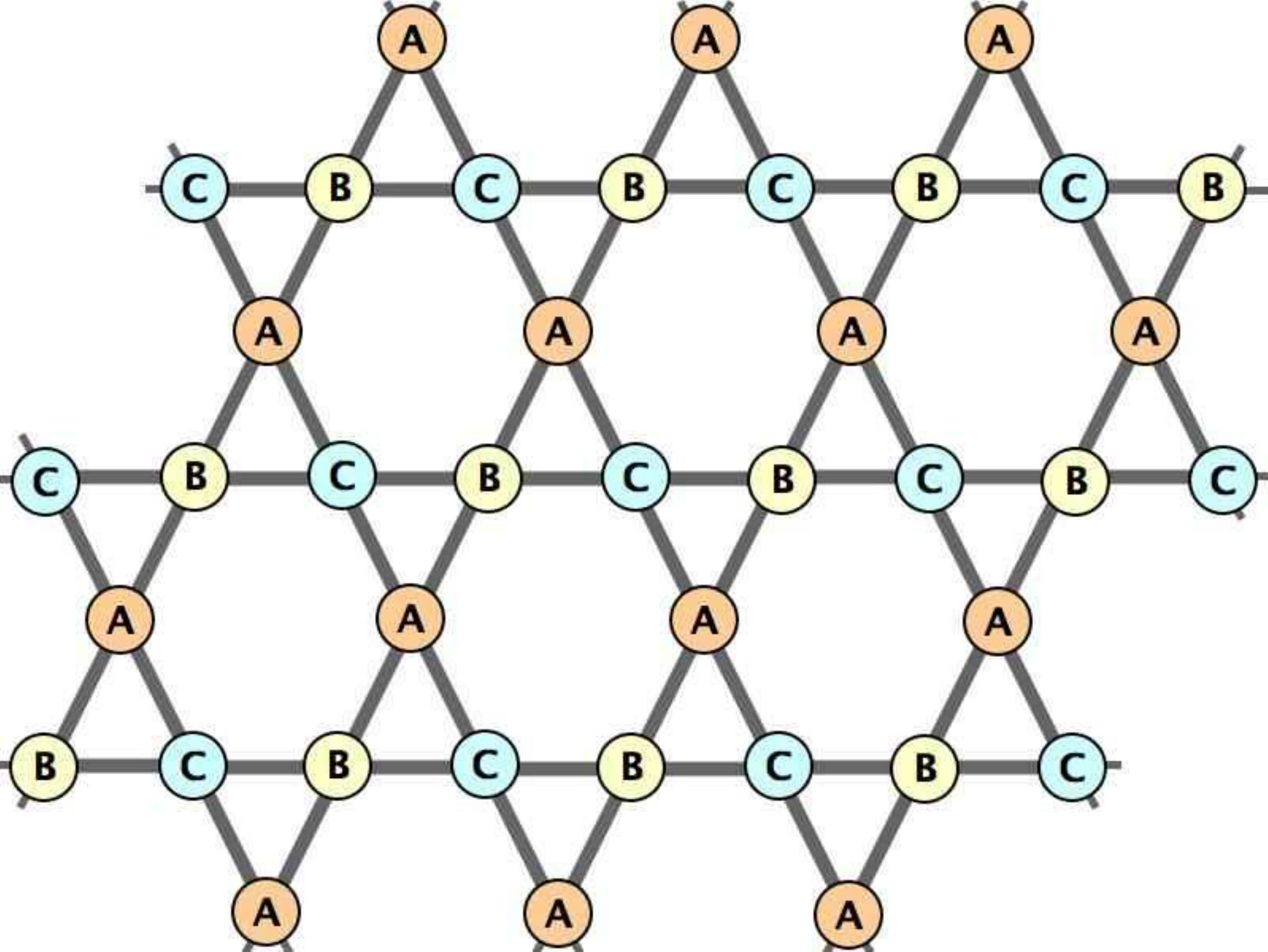}
\centering
\caption{The simplest ground state for the kagome structure. A, B, and C represent a set of mutually orthogonal ${\rm CP}^2$ vectors. }
\label{kagomeQ0}
\end{figure}

 We performed two main tests of the Monte Carlo code. The first (standard) test was to reproduce well-known results: specifically,  we recovered the critical temperature   $T\ns_\Rc\simeq0.69$ of the  classical cubic lattice  ${\rm O} (3)$ Heisenberg model \cite{Chen1993monte}. Our second concern is more unusual: namely, whether the  Metropolis algorithm is sufficiently ergodic  to generate a phase transition for a classical system governed by the unusual interaction relevant to simplex solid models: for instance, for a three-site simplex $(ijk)$ we have  
the  interaction 
$u\ns_{ijk}=-2\ln V\ns_{ijk}$, where $V\ns_{ijk}=|\eps^{\mu\nu\lambda} z\nd_{i,\mu}\,z\nd_{j,\nu}\,z\nd_{k,\lambda}|$,
is the internal volume of the triple $(ijk)$.
In order to ensure that the {\it absence} of  a transition on a more complicated lattice is not simply an artefact of our simulations, it is important to verify that such an interaction can indeed lead to a phase transition in a simple model system. To that end, we investigated a simple ${\rm SU}(3)$-invariant model on a simple cubic lattice, with
\begin{equation}
H = -2\sum_\BR \sum_{\mu=1}^3 \ln V(\BR-\ehat\ns_\mu,\BR,\BR+\ehat\ns_\mu) \quad.
\end{equation}
As this is an unfrustrated lattice, with a finite set of broken-symmetry global energy minima (up to global ${\rm SU}(3)$ rotations) and in three dimensions where fluctuation effects should not destabilize order, it is reasonable to expect a finite-temperature transition in this model. Indeed, we find a transition at $T \simeq 1.25$ or $M \simeq 0.8$, visible in both heat capacity and structure factor
calculations. Armed with this reassuring result, we now turn our attention to several specific examples in two and three dimensions.

\begin{figure}[t]
\includegraphics[width=0.8\linewidth]{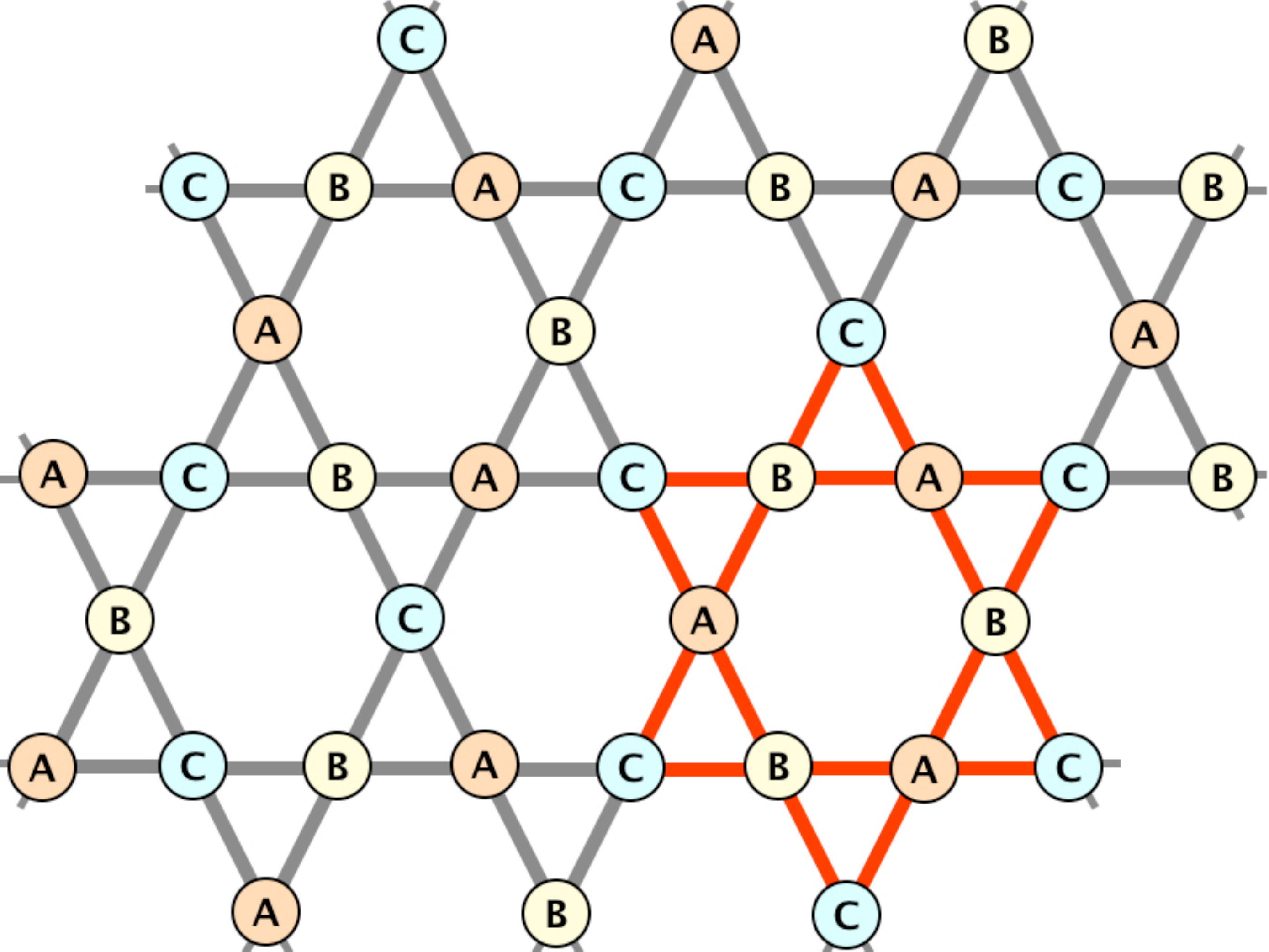}
\caption{The $\sqrt{3}\times\sqrt{3}$ kagome ground state supports an extensive number of zero-energy fluctuation modes.
A, B, and C represent a set of mutually orthogonal ${\rm CP}^2$ vectors. The red Star of David unit is used to analyze local zero modes.}
\label{kagomeSqrt3}
\end{figure}

\section{${\rm SU}(3)$ simplex solid on the kagome lattice}
As our first example, we consider the ${\rm SU}(3)$ model on the kagome lattice.  The elementary simplices of this lattice are triangles, and
$H\ns_{\rm cl}$ describes a classical model of ${\rm CP}^2$ spins with three-body interactions, {\it viz.\/}
\begin{equation}
H_{cl}=-\sum_\Gamma\ln \big|   \eps^{\alpha\nd_1 \alpha\nd_2 \alpha\nd_3}\,
z_{\Gamma\nd_1,\alpha\nd_1} z_{\Gamma\nd_2,\alpha\nd_2} z_{\Gamma\nd_3,\alpha\nd_3} \big|^2\ ,
\label{HamSU(3)}
\end{equation}
where $\Gamma\nd_i$ are the vertices of the elementary triangle $\Gamma$.  The structure factor $S\ns_{ij}(\Bk)$ is
then a $3\times 3$ matrix-valued function of $\Bk$.

In any ground state, each triangle is fully satisfied, with $|R\nd_\Gamma|=1$.  One such ground state is the so-called
$\Bq=0$ structure, which is a Potts state with
\begin{equation*}
\omega\nd_\ssr{A}=\begin{pmatrix} 1 \\ 0 \\ 0 \end{pmatrix}\quad,\quad
\omega\nd_\ssr{B}=\begin{pmatrix} 0 \\ 1 \\ 0 \end{pmatrix}\quad,\quad
\omega\nd_\ssr{C}=\begin{pmatrix} 0 \\ 0 \\ 1 \end{pmatrix}
\end{equation*}
assigned to each of the three sublattices of the tripartite kagome structure. The structure factor is
given by
\begin{equation}
S\ns_{ij}(\Bk)=\big(\delta\ns_{ij}-\frac{1}{3}\big)\cdot\Omega\,\delta\ns_{\Bk,0}\quad.
\end{equation}

Another Potts ground state is the $\sqrt{3}\times\sqrt{3}$ structure, depicted in Fig. \ref{kagomeSqrt3},
which has a nine site unit cell consisting of three elementary triangles. The structure factor is then
\begin{equation}
S\ns_{ij}(\Bk)=
{\Omega\over 3}\begin{pmatrix} 1 & \omega^2 & \omega \\ \omega & 1 & \omega^2 \\ \omega^2 & \omega & 1 \end{pmatrix}
\delta\ns_{\Bk,\BK} + 
{\Omega\over 3}\begin{pmatrix} 1 & \omega & \omega^2 \\ \omega^2 & 1 & \omega \\ \omega & \omega^2 & 1 \end{pmatrix}
\delta\ns_{\Bk,\BK'}  
\end{equation}
where $\omega=e^{2\pi i/3}$ and $\BK$ and $\BK'$ are the two inequivalent Brillouin zone corners.

We emphasize that the Potts states do not exhaust all possible ground states, because for some spin configurations,
certain collective local spin rotations are possible without changing the total energy.  The number of such zero modes can
even be extensive \cite{MoessnerChalker}.  In the case of the ${\rm SU}(4)$ model on the cubic lattice, to be discussed below,
there are only finitely many soft modes, and we observe a finite temperature phase transition.

Consider now the zero-energy fluctuations for the $\Bq=0$ structure. Six of them are global ${\rm SU}(3)$ rotations, while the
others may be constructed as follows.  Identify A, B, and C spin sublattices by different colors. There are three types of
dual-colored lines in this structure (see Fig. \ref{kagomeQ0}): ABAB, BCBC, and CACA.  The spins along each of these lines may
be rotated independently around $\omega\ns_\sigma$ axis corresponding to the third color. This is a source of zero modes:
each line provides two zero modes, but total number of zero modes in this structure is still sub-extensive, scaling as
$\Omega^{1/2}$.  

\begin{figure}[t]
  \centering
\includegraphics[width=1.0\linewidth]{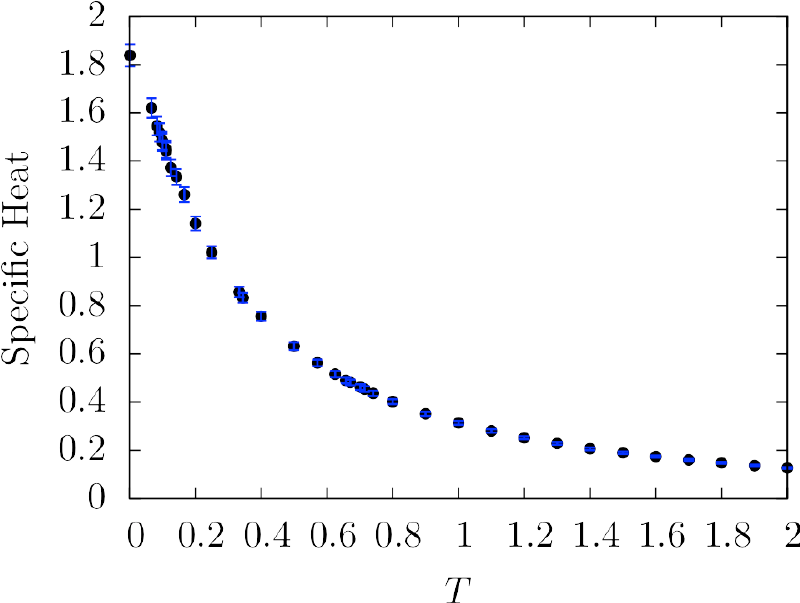}
\caption{Specific heat per site {\it versus\/} temperature for the kagome structure with $N=1296$ sites.}
\label{heat2D}
\end{figure}

\begin{figure}[t]
\centering
\includegraphics[width=1.0\linewidth]{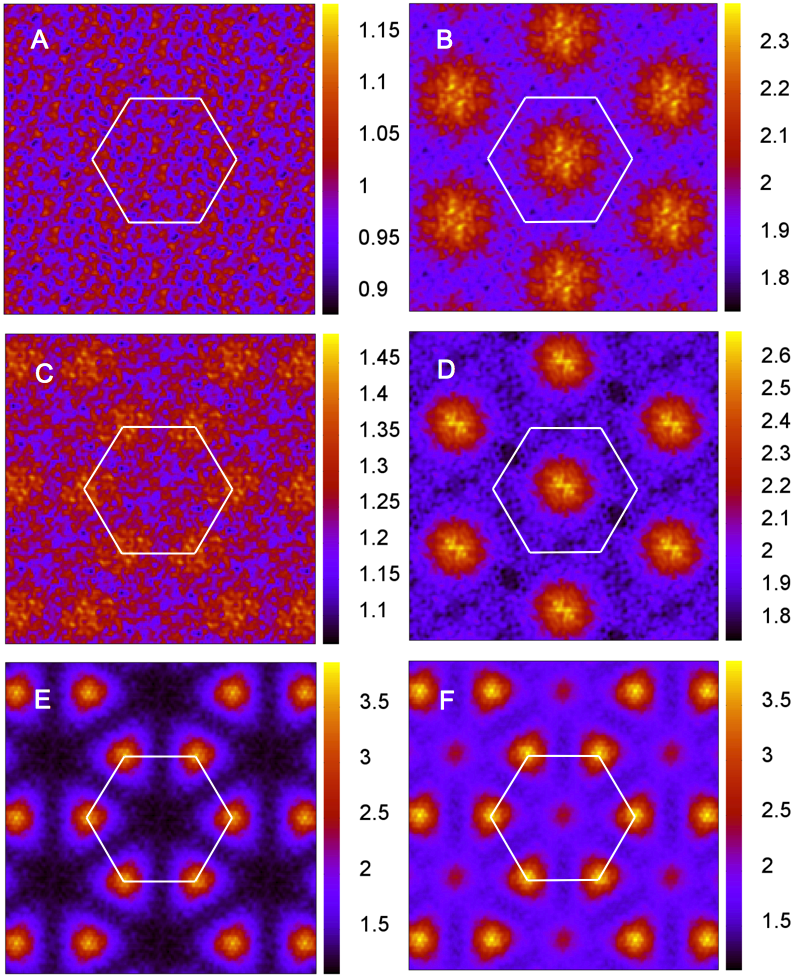}\hfill
\caption{SU(3) kagome lattice model ($N=1296$ sites).  Left (A,C,E): largest eigenvalue, Right (B,D,F): sum of all eigenvalues.
White hexagon indicates the border of Brillouin zone. A,B correspond to high temperature ($M=2$); C,D to intermediate temperature ($M=5$);
and  E,F to low temperature ($M=20$).}
\label{highT2D}
\end{figure}

For the $\sqrt{3}\times\sqrt{3}$ structure of Fig. \ref{kagomeSqrt3}, there is an extensive set of zero modes.  Consider
the case of a single Star of David from this structure, depicted in red in the figure.
The internal hexagon is a six-site loop surrounded by six external spins. If the loop spins belong to the plane spanned by
vectors $z_\ssr{A}$ and $z_\ssr{B}$ while the external spins are all $z_\ssr{C}$, there is a local zero-energy mode associated
with the hexagon which rotates $z_\ssr{A}$ and $z_\ssr{B}$ about $z_\ssr{C}$, while keeping all three spins mutually orthogonal.
For a single six-site loop with six additional vertices this type of fluctuation coincides with the global rotation, but in the
lattice we can rotate each of the loops independently. This leads to the extensive number of zero modes, which increases the
entropy. Fluctuations about the Potts state yield a heat capacity of $C=\frac{16}{9} \approx 1.78$ per site. The counting of modes
is as follows.  There are four quadratic modes per site.  Any individual hexagon, however, can be rotated by a local U(2) matrix in
the subspace perpendicular to the direction set by its surrounding spins ({\it e.g.\/}, an AB hexagon can be rotated about the C
direction). There are two independent real variables associated with such a rotation. (For an AB hexagon, the A sites are orthogonal
to the C direction, hence $z\nd_\ssr{A}$ is specified by two complex numbers, plus the constraint of $z\yd_\ssr{A}z\nd_\ssr{A}=1$
and the equivalence under $z\nd_\ssr{A}\to e^{i\alpha}z\nd_\ssr{A}$.)  Subtracting out the zero modes, we find the heat capacity per
site would then be $C(0)=\half\times\big(4-\frac{2}{3}\big)=\frac{5}{3}$.  However, we have subtracted too much.  Only one third
of the hexagons support independent zero modes (the AB hexagons, say).  The remaining two thirds are not independent and will contribute
at quartic order in the energy expansion.  The specific heat contribution from these quartic modes is then
$\frac{1}{4}\times\frac{2}{3}\times\frac{2}{3}=\frac{1}{9}$.  Thus, we expect $C(0)=\frac{16}{9}$.  This analysis of the 
zero modes in both structures follows that for the ${\rm O}(3)$ Heisenberg antiferromagnet on the kagome lattice
\cite{CHS92}.  As in the ${\rm O}(3)$ case, the low temperature entropy selects configurations which are locally close
to the $\sqrt{3}\times\sqrt{3}$ structure.  This order by disorder\footnote{OBD1980} (OBD) mechanism was shown in Ref.~\onlinecite{arovas:104404} by
invoking a global length constraint which turns the low temperature Hamiltonian of eqn. \ref{HLT} into a spherical model,
introducing a single Lagrange multiplier $\lambda$ to enforce $|\chi|^2+{1\over \Omega}\sum_i\langle \pi\yd_i\pi\nd_i\rangle=1$,
where $\chi$ plays the role of a condensate amplitude.  The free energy per site is then
\begin{equation}
f=-\lambda+\lambda|\chi|^2 + (N-1)\,T\!\int\limits_0^\infty\!\! d\ve\>g(\ve)\ \ln\!\bigg({\ve+\lambda\over T}\bigg)\quad.
\label{ferg}
\end{equation}
Extremizing with respect to $\lambda$ yields the saddle point equation, and the OBD selection follows from a consideration of saddle-point free energies of the $\Bq=0$ and $\sqrt{3}\times\sqrt{3}$ states.

We now turn to the results of our Monte Carlo simulations. The heat capacity $C(T)$ per site is shown in Fig. \ref{heat2D}.  We find $C(T)$ exhibits no singularities at any finite
temperature and remains finite at zero temperature.  Thus, there is no phase transition down to $T=0$.  Note that while the Hohenberg-Mermin-Wagner theorem forbids the breaking of the continuous ${\rm SU}(3)$ symmetry at finite temperatures (since the classical Hamiltonian $H\ns_{\rm cl}$ is that of a two-dimensional system
with finite-range interactions) it leaves open the possibility of a transition due to breaking a discrete lattice symmetry. That such a transition does not occur -- as evinced by the absence of any specific heat singularities -- is a nontrivial result of these simulations. From equipartition, we should expect $C=2$ if all freedoms appear quadratically in the
effective low energy Hamiltonian. Instead, we find $C(0)=1.84 \pm 0.03$. The fact that the heat capacity is significantly
lower than $2$ suggests that there is an extensive number of zero modes or or other soft modes.

Although the absence of any phase transition in the specific heat data suggests that there is no true long-range order in the kagome system even at $T=0$, it leaves open the question of whether there is some form of incipient local order in the system as $T\rightarrow 0$.
To further investigate the local order at low temperatures, we turn to the structure factor $S\ns_{ij}(\Bk)$.  Recall that this is a
$3\times 3$ matrix for the kagome lattice, and we have focused our attention on the eigenvalue of maximum amplitude
as well as the trace of this matrix.  Our Monte Carlo results for these quantities are plotted in Fig. \ref{highT2D}.
At high temperatures, we find the only detectable structure has the same periodicity as the lattice, with ${\rm Tr}\,S(\Bk)$
exhibiting a peak at the center of the Brillouin zone. Upon lowering temperature, one can see that additional structure emerges,
and the peak shifts to the Brillouin zone corners $\BK$ and $\BK'$, corresponding to the $\sqrt{3} \times \sqrt{3}$ structure.
The width of the structure factor peaks remains finite down to $T=0$, and there are no true Bragg peaks.
The heat capacity of the ideal $\sqrt{3} \times \sqrt{3}$ structure is somewhat lower than
the heat capacity obtained from Monte Carlo simulations.

\begin{figure}[t]
\centering
\includegraphics[width=1.0\linewidth]{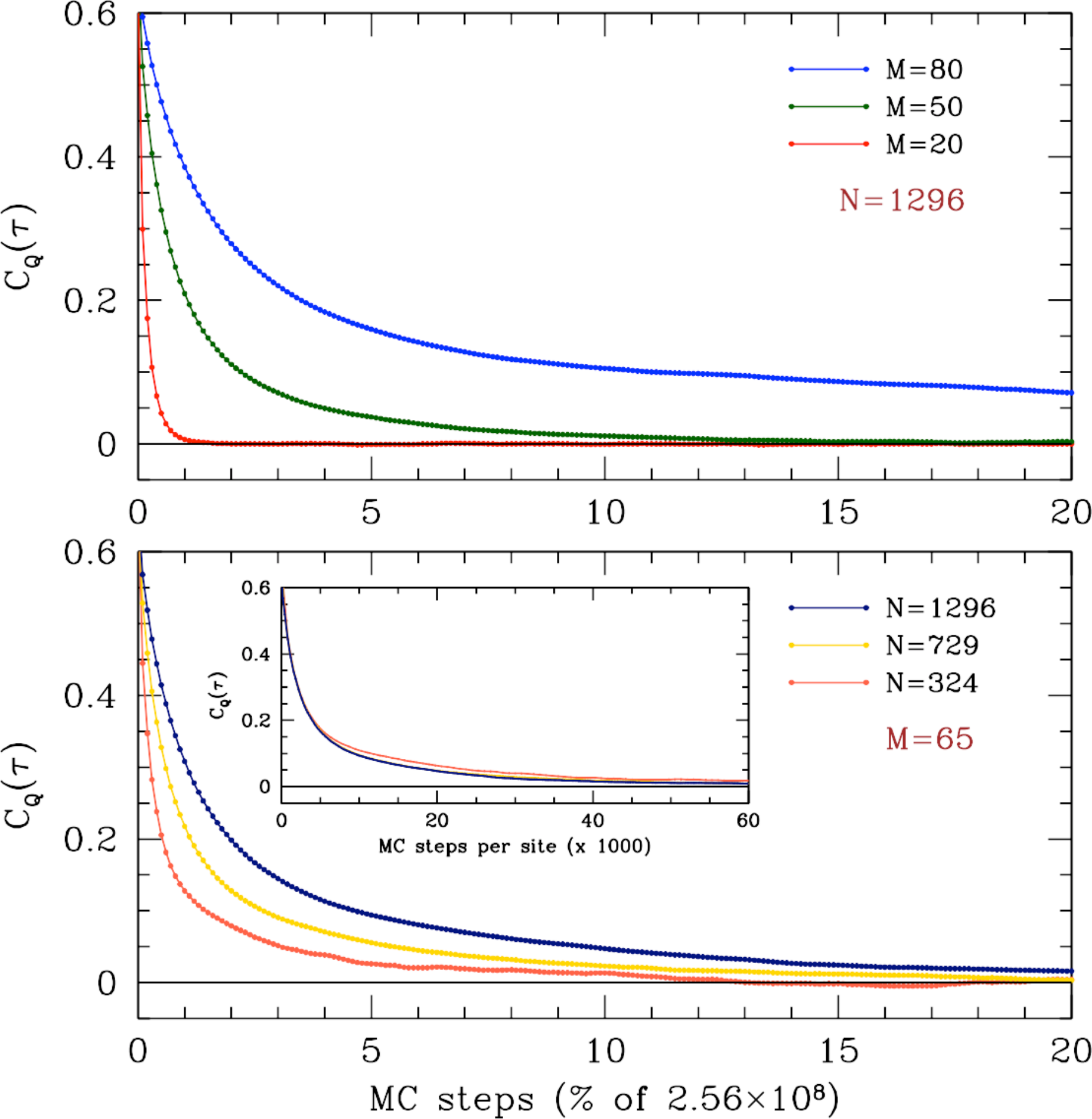}
\caption{Autocorrelation function $C_Q(\tau)$ {\it versus\/} Monte Carlo time for the SU(3) model on the kagome lattice.  Upper panel:  Behavior for $N=1296$ site system
at inverse temperatures $M=20$, $M=50$, and $M=80$. Bottom panel: Behavior for $M=65$ data for different sized systems.  Inset shows $C_Q(\tau)$ {\it versus\/} Monte Carlo
steps per site. The overlap for different $N$ values indicates size-independence of the results.}
\label{2Dauto}
\end{figure}

Further insight on the nature of the low-temperature state of the kagome simplex solid is afforded by studying the autocorrelation function
$C_Q(\tau' - \tau) = \langle {\rm Tr}\,\big[Q(i,\tau)\,Q(i,\tau')\big]\rangle $, 
where additional averaging was performed over the starting time $\tau$ and the site index $i$.
As is clear from Fig. \ref{2Dauto}, the autocorrelator vanishes for $|\tau'-\tau|\to\infty$, consistent with a lack of long-range order.  For larger $M$ (smaller $T$),
the dynamics slow down, consistent with the dominance of the local $\sqrt{3}\times\sqrt{3}$ pattern in the low-temperature structure factor.

\section{Three-dimensional lattices}
\subsection{${\rm SU}(3)$ simplex solid on the hyperkagome lattice}
\begin{figure}[t]
\includegraphics[width=0.75\linewidth]{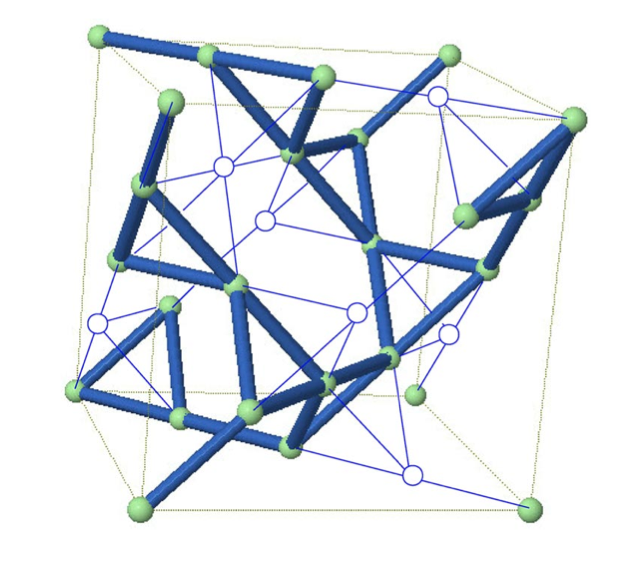}
\caption{The hyperkagome structure (from Ref.~\onlinecite{Hopkinson99}).}
\label{fig:hyperK}
\end{figure}

We embark on our analysis of three-dimensional lattices by considering the analog of the kagome in three dimensions: the imaginatively-named hyperkagome lattice (Fig.~\ref{fig:hyperK}). This is a three-dimensional fourfold coordinated lattice consisting of loosely-connected triangles. The
crystal structure is simple cubic, with a 12-site basis.  It may be described as a depleted pyrochlore structure, where one
site per pyrochlore tetrahedron is removed.  With triangular simplices, we again have the Hamiltonian of eqn. \ref{HamSU(3)}, but here owing to the increased dimensionality, we might expect that ordered states remain relatively stable to fluctuation effects.

There is a vast number of ground states of the SU(3) simplex solid model on the hyperkagome lattice. We first consider the
simplest ones, Potts states, where three mutually orthogonal CP$^2$ vectors $\omega\nd_\ssr{A,B,C}$ are assigned to the lattice
sites such that the resulting arrangement is a ground state, where the volume of each triangle $(ijk)$,
$|R\nd_\Gamma|=|\eps^{\alpha\beta\gamma}z\nd_{i,\alpha}z\nd_{j,\beta}z\nd_{k,\gamma}|$ is maximized, {\it i.e.\/}
$|R\nd_\Gamma|=1$. 

\begin{figure}[b]
\includegraphics[width=1.0\linewidth]{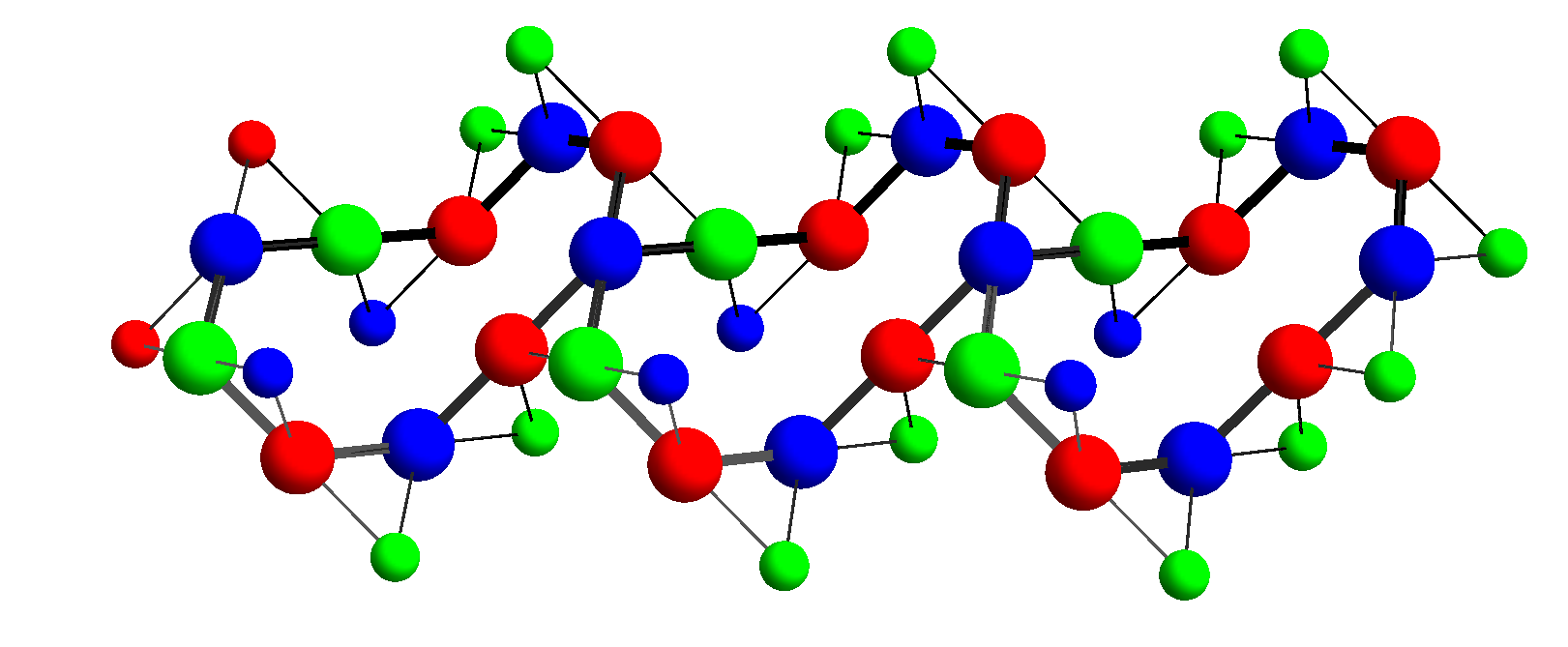}
\includegraphics[width=1.0\linewidth]{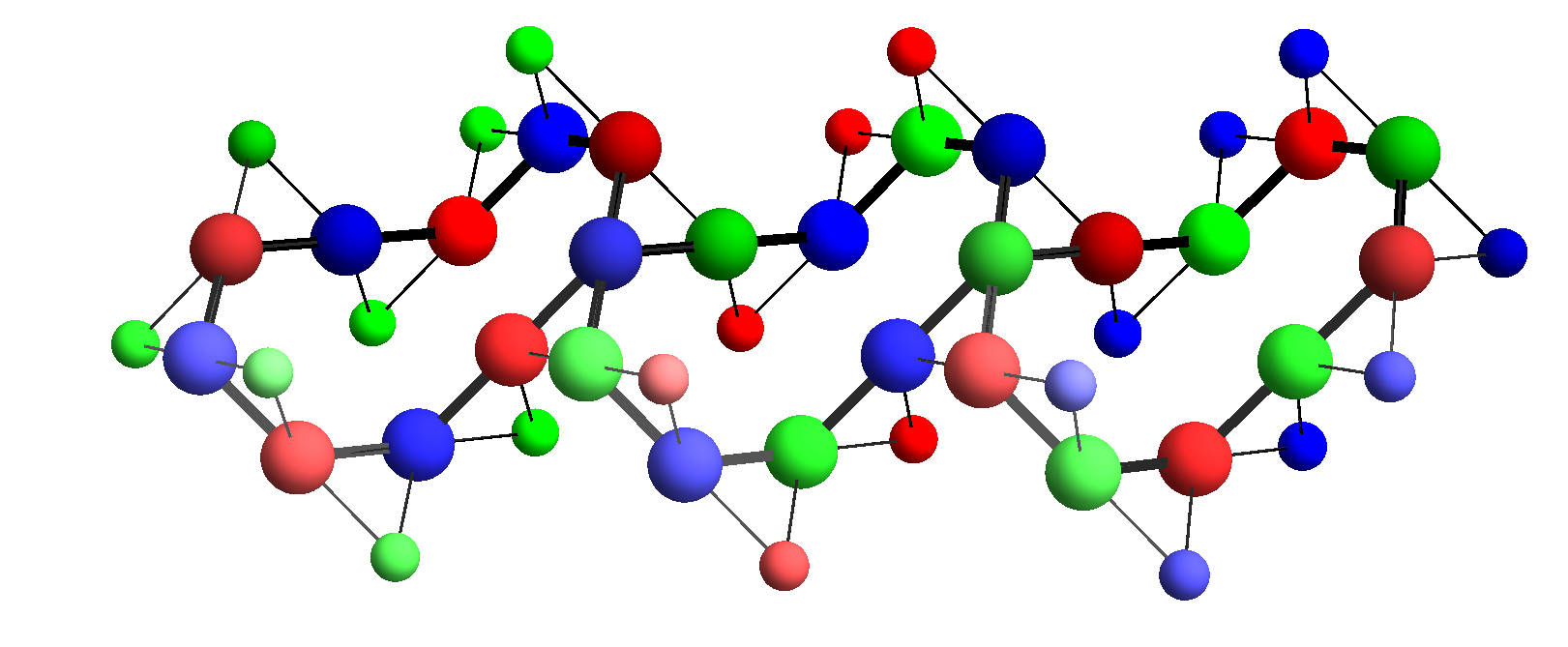}
\caption{Top: Three unit cells, 12 sites each, of the $\Bq=0$ structure. The 10-site loops
do not support any zero modes.  Bottom: Unit cell consisting of 36 sites of the structure analogous to $\sqrt{3} \times \sqrt{3}$ in the case of kagome lattice. Thick lines indicate 10-site loops which provide zero modes. The leftmost 10-site red-blue loop can be rotated about the green direction, yielding a zero mode. In both panels, sites on the loops are shown with large spheres,
and neighboring off-loop sites with small spheres.}
\label{fig:hyperkagome_q0q3}
\end{figure}

The simplest Potts ground state will have the same periodicity as the lattice ($\Bq=0$), with its 12 site unit cell.  Computer
enumeration reveals that there are two inequivalent $\Bq=0$ structures, one of which is depicted in the top panel of
Fig. \ref{fig:hyperkagome_q0q3}. Potts ground states with larger unit cells are also
possible, and an example of a Potts state with a 36 site unit cell is shown in the bottom panel of the figure.
Such structures are analogs of $\sqrt{3} \times \sqrt{3}$ structure on the kagome lattice, discussed in the previous section.

\begin{figure}[t]
\includegraphics[width=1.0\linewidth]{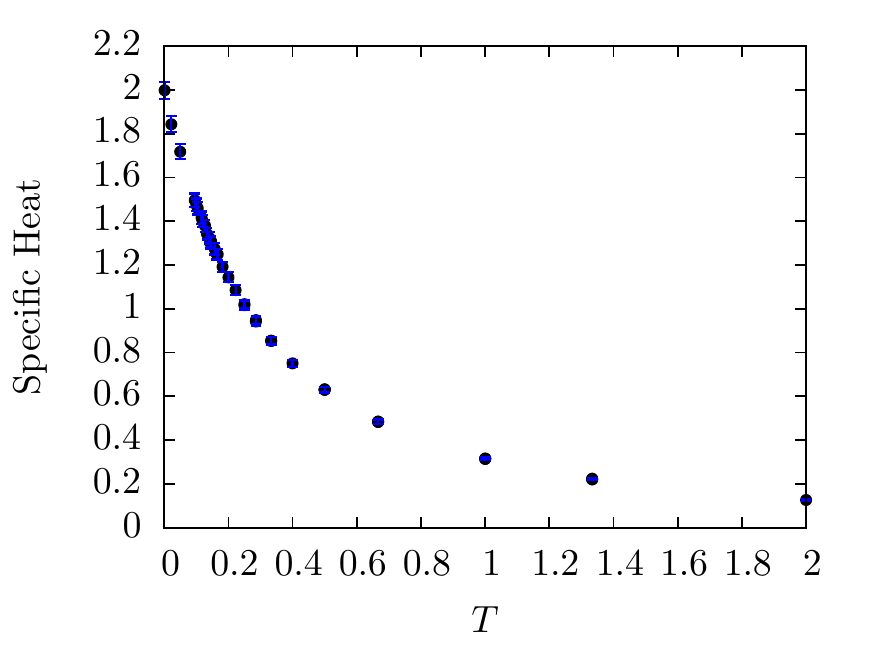}
\caption{Specific heat for the SU(3) model on the hyperkagome lattice with $N=2592$ sites.}
\label{fig:heat_3D}
\end{figure}

\begin{figure}[b]
\centering
\includegraphics[width=1.0\linewidth]{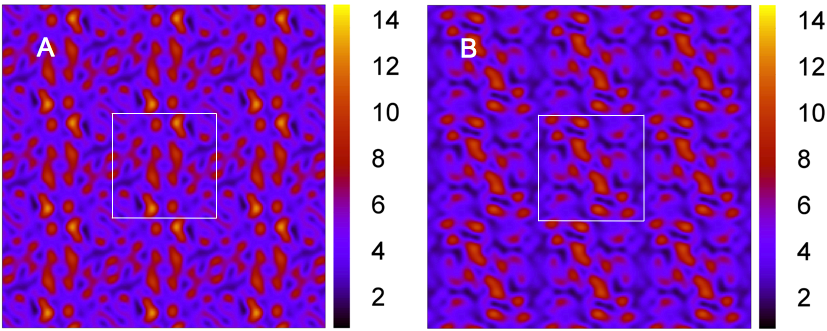}
\caption{Structure factor for the SU(3) hyperkagome model at $T=0.01$ ($M=100$).  Results show $S(\Bk)$ in the $(k_x,k_y)$
plane with $k_z = 0$ (A) and $k_z = \pi / a$ (B) . White lines denote the borders of the Brillouin zone. Number of sites is 6144 ($8^3$ unit cells).}
\label{hyperkagome_factors}
\end{figure}

Monte Carlo simulations of $H\nd_{\rm cl}$ on the hyperkagome lattice show no cusp in $C(T)$, suggesting $T\nd_{\rm c}=0$
(Fig. \ref{fig:heat_3D}). In contrast to the kagome, structure factor measurements exhibit a diffuse pattern spread throughout the Brillouin zone and are insufficient to show which low temperature structure is preferred (Fig. \ref{hyperkagome_factors}). 

The six-site loops in the 2D kagome lattice have an analog in the 3D hyperkagome structure, which contains ten-site loops.
For the (2D) kagome model, the six-site loops support zero modes in the $\sqrt{3}\times\sqrt{3}$ Potts state.  There is an
analog of this degeneracy in the (3D) hyperkagome model, where the corresponding Potts state features a 36-site unit cell,
mentioned above and depicted in Fig. \ref{fig:hyperkagome_q0q3}.
The zero mode corresponds to a SU(3) rotation of all CP$^2$ spins along a 10-site loop, about a common axis.
This is possible because all the spins along the loop lie in a common CP$^2$ plane, forming an ABAB$\,\cdots$ Potts configuration.  A computer enumeration finds that there are 12 distinct such 10-site loops associated with each (12-site)
unit cell. If the hyperkagome  emulates the kagome, we expect that owing to the abundance of zero modes, structures with such loops will dominate the low-temperature dynamics of $H_{\text{cl}}$.

\begin{table*}[t]
\begin{center}
\begin{tabular}{|c|c|c|c|c|c|}
\hline
SU(3) system & A & B & C & D & E \\ \hline
$p_{\bullet\bullet\bullet}$ & $0.377 \pm 0.004$ & $0.364 \pm 0.007$ & $0.0093 \pm 0.0006$ & $0.516 \pm 0.004$ & $0.488 \pm 0.006$ \\ \hline
$p_{\bullet\bullet\circ\bullet}$ & $0.2415 \pm 0.0008$ & $0.253 \pm 0.002$ & $0.0093 \pm 0.0006$ & $0.259 \pm 0.003$ & $0.253 \pm 0.001$ \\ \hline
$p_{\bullet\bullet\circ\circ\bullet}$ & $0.305 \pm 0.002$ & $0.339 \pm 0.003$ & $0.0093 \pm 0.0003$ & $0.355 \pm 0.007$ & $0.369 \pm 0.004$ \\ \hline 
$p_{\bullet\circ\bullet\circ\bullet}$ & $0.063 \pm 0.002$ & $0.043 \pm 0.003$ & $0.00025\pm 0.00001$ & $0.138 \pm 0.003$ & $0.127 \pm 0.005$ \\ \hline
$\lambda_{\rm min}$ & $-0.167  \pm 0.001$ & $-0.150  \pm 0.002$ & $-0.3296 \pm 0.0001$ & $-0.088 \pm 0.002$ & $-0.110 \pm 0.004$ \\ \hline
$E/N_{\triangle}$ & $0.02945 \pm 0.00001$ & $0.02945 \pm 0.00001$ & $0.01885 \pm 0.00002$ & $0.01984 \pm 0.00001$ & $0.029518 \pm 0.000003$\\ \hline 
\end{tabular}
\end{center}
\caption{10 site loop statistics in the SU(3) hyperkagome model (see text). A) hyperkagome (lowest $\lambda_{\rm min}$). B) hyperkagome
(lowest $p_{\bullet\circ\bullet\circ\bullet}$).  C) 10 sites (uniform boundary).  D) 10 sites (no zero mode).
E) 20 sites (loop + boundary).  The inverse temperature is $M=100$.}
\label{table_hyper}
\end{table*}

In order to characterize the structure revealed by our Monte Carlo simulations, it is convenient to first define a series of `loop statistics' measures that serve as proxies for the local correlations of the spins.
As before, we define the volume for the triple of sites $(i,j,k)$ as
\begin{equation}
V(i,j,k)=\big|\eps^{\mu\nu\lambda} z\nd_{i,\mu}\,z\nd_{j,\nu}\,z\nd_{k,\lambda}\big|\quad,
\end{equation}
{\it i.e.\/} $V(i,j,k)=|R\nd_\Gamma|$ (see eqn. \ref{RGamma}), where $\Gamma$ denotes a triangle with vertices $(i,j,k)$. The value of $V^2(i,j,k)$ for different choices of triples in a ten-site loop will serve as our primary statistical measure. 
Note that $0\le V(i,j,k)\le 1$, with $V=0$ if any two of the CP$^2$ vectors $\{z\nd_i,z\nd_j,z\nd_k\}$ are parallel, and
$V=1$ if they are all mutually perpendicular.  If the CP$^2$ vectors were completely random from site to site,
then the average over three distinct sites would be $\big\langle V^2(i,j,k)\big\rangle)=\frac{2}{9}$. For an
ABAB$\,\cdots$ Potts configuration, $V(i,j,k)=0$ for any three sites along the loop. We then define the loop statistics measures
\begin{align}
p\nd_{\bullet\bullet\bullet}&=\Big\langle\frac{1}{10}\sum_{i=1}^{10}V^2(i,i+1,i+2)\Big\rangle\\
p\nd_{\bullet\bullet\circ\bullet}&=\Big\langle\frac{1}{10}\sum_{i=1}^{10}V^2(i,i+1,i+3)\Big\rangle\\
p\nd_{\bullet\bullet\circ\circ\bullet}&=\Big\langle\frac{1}{10}\sum_{i=1}^{10}V^2(i,i+1,i+4)\Big\rangle\\
p\nd_{\bullet\circ\bullet\circ\bullet}&=\Big\langle\frac{1}{10}\sum_{i=1}^{10}V^2(i,i+2,i+4)\Big\rangle\quad,
\end{align}
where the angular brackets denote thermal averages and averages over unit cells. 

\begin{figure}[b]
\centering
\includegraphics[width=0.5\linewidth]{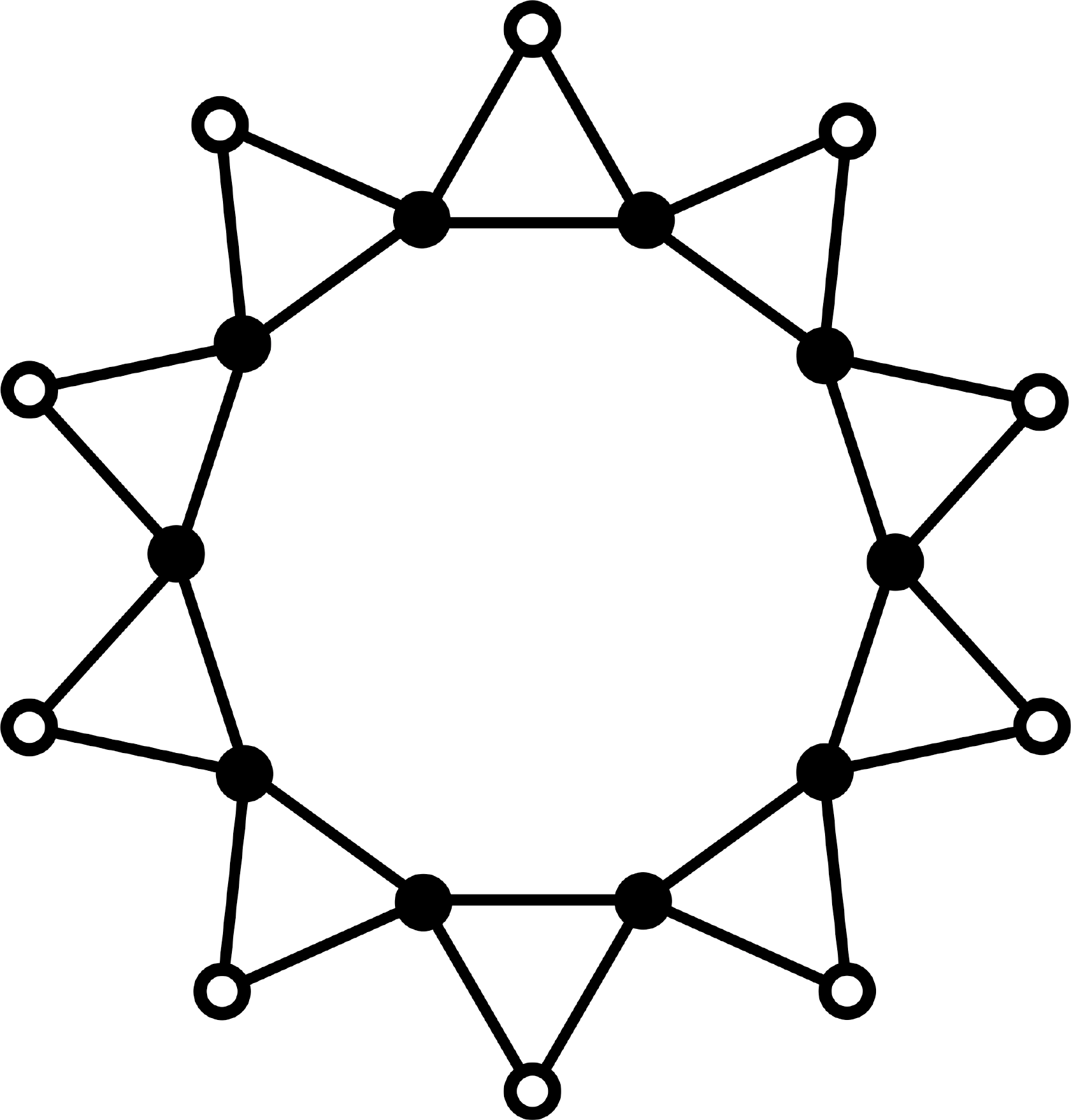}
\caption{A ten site loop surrounded by ten boundary sites.}
\label{fig_badge}
\end{figure}

Another useful diagnostic is
to compute the eigenspectrum of the gauge-invariant tensor $Q\nd_{\mu\nu}(i)$ averaged over sites,
\begin{equation}
{\overline Q}\nd_{\mu\nu}\equiv \frac{1}{10}\!\!\sum_{i\in{\rm loop}}\!\! \langle z^*_{i,\mu}\,z\nd_{i,\nu}\rangle - \
\frac{1}{3} \delta\nd_{\mu\nu}\quad.
\end{equation}
For randomly distributed CP$^2$ vectors, ${\overline Q}=0$.  If the loop is in the ABAB$\cdots$ Potts configuration,
${\overline Q}=\frac{1}{6}-\frac{1}{2} P\nd_\ssr{C}$, where $P\nd_\ssr{C}$ is the projector onto the C state
orthogonal to both A and B.   Our final diagnostic is the average energy per triangle, denoted $E/N_{\triangle}$.

Statistical data for the 10-site loops at inverse temperature $M=100$ are shown in Table \ref{table_hyper}, where
four structures are compared. Each column of the table refers to a particular class of 10 site loop.  The first two columns
present Monte Carlo data for a 6144 site lattice ($8^3$ unit cells) with periodic boundary conditions. Averages are performed
over the entire lattice. In the column A, the particular loop among the 12 distinct representatives per unit cell is chosen on the basis
of the lowest eigenvalue of ${\overline Q}\nd_{\mu\nu}$. In column B, the representative loop has the lowest value of
$p\nd_{\bullet\circ\bullet\circ\bullet}$.  In column C, data from a single 10-site loop with a fixed set
of boundary spins, as depicted in Fig. \ref{fig_badge}, is presented.  In this case the boundary spins are all parallel
CP$^2$ vectors, hence for $T=0$ the ground state of this ring would be a Potts state of the ABAB$\cdots$ type, and indeed the
data are close to what we would predict for such a Potts state, where the internal volume $V(i,j,k)$ vanishes for any
triple of sites on the loop, and where the eigenvalues of ${\overline Q}$ are
$\big\{\!-\!\frac{1}{3},\frac{1}{6},\frac{1}{6}\big\}$.
Such a configuration exhibits a zero mode, since the loop spins can be continuously rotated about the direction set by the
boundary.  If we fix the boundary spins such that there is no such zero mode, and average over all such boundary configurations,
we obtain the data in column D.  Finally, column E presents data for the 20-site system shown in Fig. \ref{fig_badge}, where
the boundary spins are also regarded as free.

\begin{figure}[b]
\centering
\includegraphics[width=0.85\linewidth]{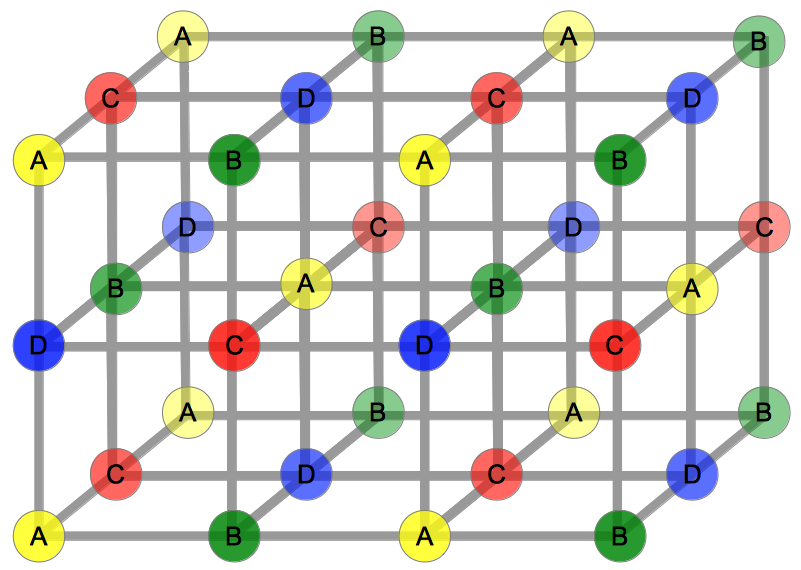}
\caption{Potts ground state of {\rm SU}(4) classical model on a cubic lattice has a bcc structure.}
\label{bcc_cubic}
\end{figure}

Our results lead us to conclude that the SU(3) model on the hyperkagome lattice is unlike the planar kagome case in that there it is
far from a Potts state, even at low temperatures.  There is no thermodynamically significant number of ABAB$\cdots$ ten-site loops,
and the statistics of these loops in the hyperkagome structure
most closely resemble the results in the last column of Tab. \ref{table_hyper},
corresponding to a single loop with a fluctuating boundary.  This is supported by static structure factor data in Fig.
\ref{hyperkagome_factors}, which shows no discernible peaks.  In addition, the heat capacity, shown in Fig. \ref{fig:heat_3D},
tends to the full value of $C(T=0)=2N$, corresponding to four quadratic degrees of freedom per site.  

\subsection{${\rm SU}(4)$ model on the cubic lattice}
Thus far we have considered models with corner-sharing simplices.  We now consider a 3D model with edge-sharing simplices.  The
individual spins are four component objects lying in the space CP$^3$.  These may be combined into singlets using the plaquette
operator $\phi\yd_\Gamma=\eps^{\mu\nu\lambda\rho}\,b\yd_{i,\mu}\,b\yd_{j,\nu}\,b\yd_{k,\lambda}\,b\yd_{l,\rho}$, where $(ijkl)$
are the sites of the $4$-simplex $\Gamma$. On a cubic lattice, $M$ such singlets are placed on each elementary face, so each site is
in a fully symmetric representation of ${\rm SU}(4)$ with $12M$ boxes. Note that two faces may either share a single edge, if they
belong to the same cube, or a single site. Again with $T=1/M$, we have identified a second order phase
transition of the corresponding classical system using Monte Carlo simulation.  The classical Hamiltonian for the model is
\begin{equation}
H_{\rm cl}=-\sum_\Gamma\ln \big|   \eps^{\alpha\nd_1 \alpha\nd_2 \alpha\nd_3 \alpha\nd_4}\,
z_{\Gamma\nd_1,\alpha\nd_1} z_{\Gamma\nd_2,\alpha\nd_2} z_{\Gamma\nd_3,\alpha\nd_3} z_{\Gamma\nd_4,\alpha\nd_4} \big|^2\ ,
\label{HamSU(4)}
\end{equation}
where $\Gamma\nd_i$ are the corners of the elementary square face $\Gamma$.  An $E\ns_0=0$ ground state can be achieved by
choosing four mutually orthogonal vectors $\omega\ns_\sigma$ and arranging them in such a way that corners of every face are
different vectors from this set. The volume spanned by vectors of every simplex is then $|R\ns_\Gamma|=1$. This ground state is unique up to a global ${\rm SU}(4)$ rotation, and has a bcc structure, as shown in Fig. \ref{bcc_cubic}. Other ground states
could be obtained from the Potts state by taking a 1D chain of spins lying along one of the main axes, say ACAC, and rotating
these spins around those in the BD plane.  We see that number of zero modes is sub-extensive, however.

\begin{figure}[t]
\centering
\includegraphics[width=1.0\linewidth]{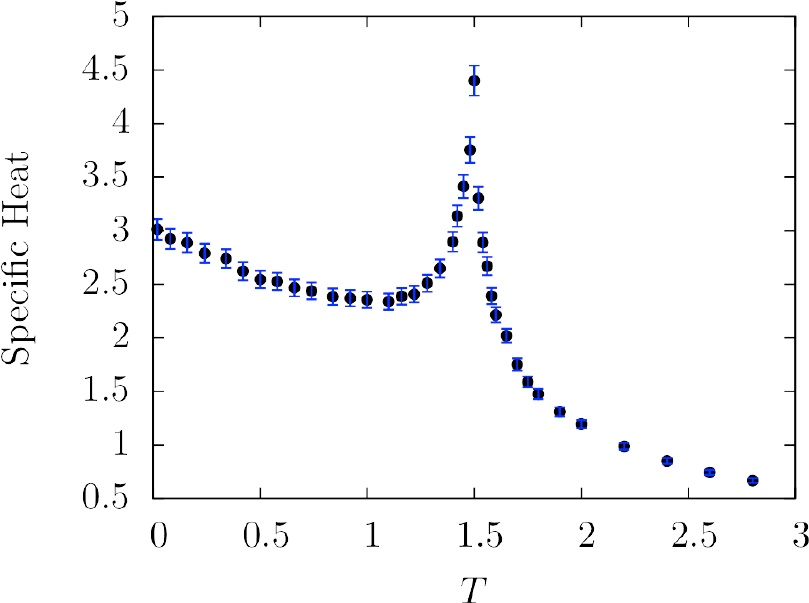}
\caption{Specific heat for ${\rm SU}(4)$ model on the cubic lattice with $N=8000$ sites.  The phase transition occurs at $T_\Rc\simeq 1.50$.}
\label{su4heat}
\end{figure}

\begin{figure}[b]
\centering
\includegraphics[width=\linewidth]{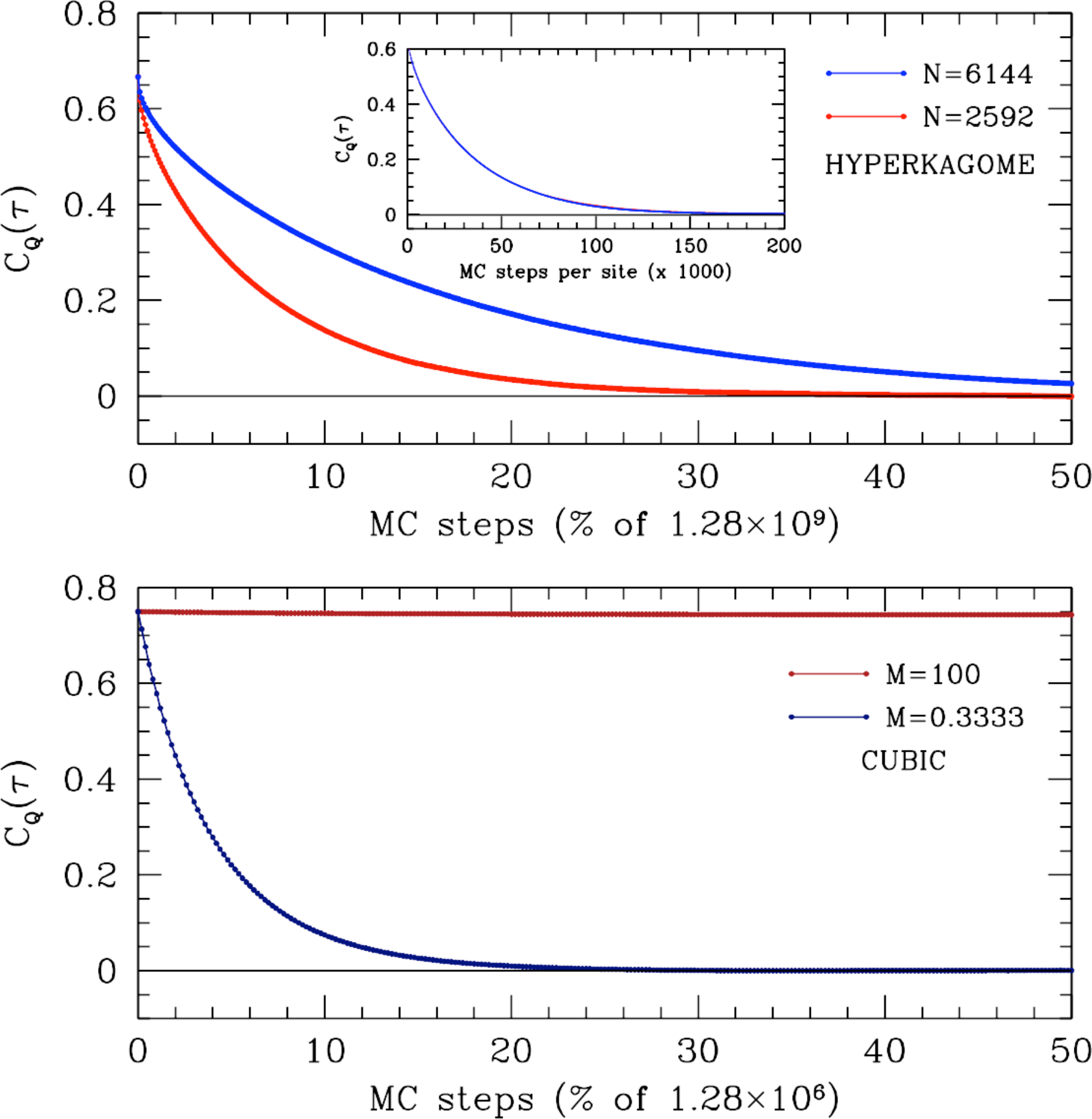}
\caption{Autocorrelation functions $C_Q(\tau)$ {\it versus\/} Monte Carlo time for two three-dimensional models.  Upper panel: hyperkagome lattice SU(3) model for $N=2592$ and
$N=6144$ sites.  Inset shows autocorrelation {\it versus\/} Monte Carlo steps per site; the data for the two sizes are overlapping.  Lower panel: cubic lattice SU(4) model.
The system is ordered at low temperature ($M=100$) and disordered at high temperature ($M=0.3333$).}
\label{su4auto}
\end{figure}

\begin{figure}[t]
\centering
\includegraphics[width=1.0\linewidth]{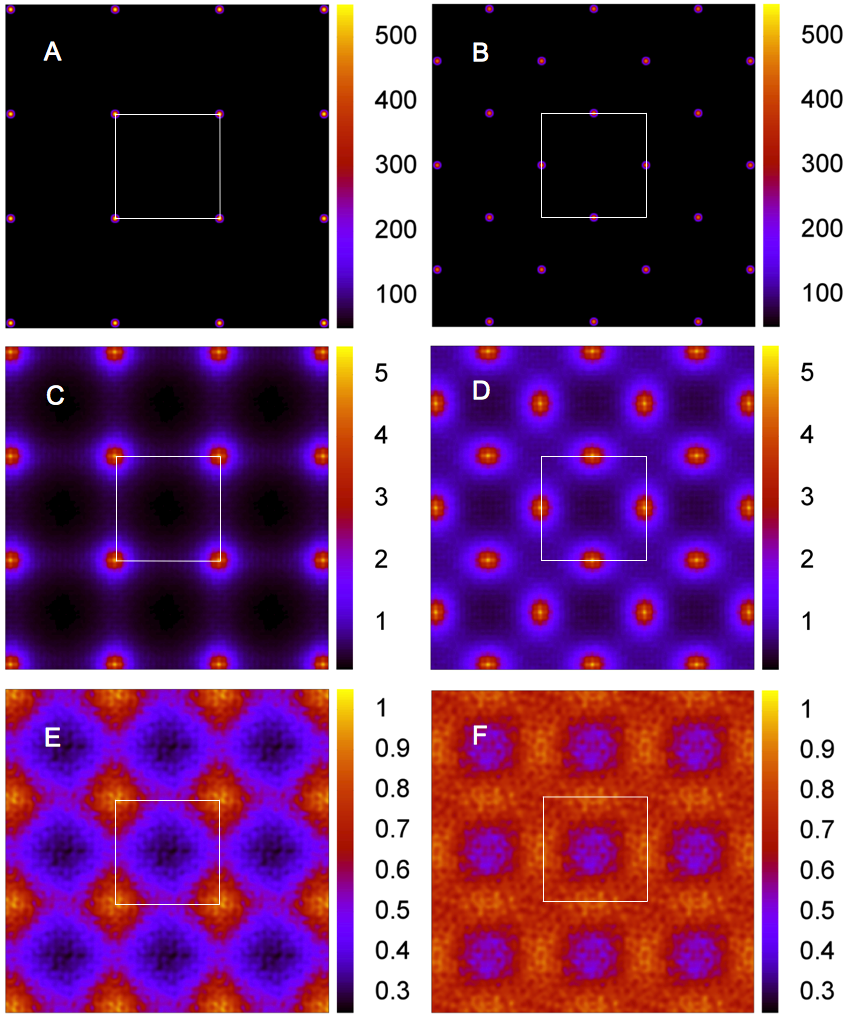}
\caption{Static structure factor for the ${\rm SU}(4)$ model on the cubic lattice ($16^3$ sites) as a function of $(k_x,k_y)$ for
$k_z=0$ (A,C,E) and $k_z=\pi$ (B,D,F) and $T={1\over 2}$ (A,B), $T=1.5$ (C,D), $T=3$ (E,F).  Note the emergence of Bragg peaks
for $T<T_\Rc\simeq 1.50$.}
\label{su4order}
\end{figure}

There is a phase transition to the ordered phase at $T = 1.485 \pm 0.005$. This is confirmed by both heat capacity
temperature dependence (Fig. \ref{su4heat}) and static factor calculations. 
Our static structure factor calculations prove the spin pattern forms a bcc lattice below the critical temperature (Fig. \ref{bcc_cubic} A-B). On the cubic lattice, $S\ns_{ij}(\Bk)=S(\Bk)$ is
a scalar, and in the Potts state of Fig. \ref{bcc_cubic} it is given by
\begin{equation}
\begin{split}
S(\Bk)&={1\over\Omega}\sum_{\BR,\BR'}{\rm Tr}\big[Q(\BR)\,Q(\BR')\big]\,e^{i\Bk\cdot(\BR-\BR')}\\
&=\frac{1}{4}\,\Omega\,\big(\delta\ns_{\Bk,\BM}+\delta\ns_{\Bk,\BM'}+\delta\ns_{\Bk,\BM''}\big)\quad,
\end{split}
\end{equation}
where $\BM=(0,\pi,\pi)$, $\BM'=(\pi,0,\pi)$, and $\BM''=(\pi,\pi,0)$ are the three inequivalent edge centers of the
Brillouin zone, resulting in an edge-centered cubic pattern in reciprocal space.
Since $T\ns_{\rm c}>1$, we have $M\ns_{\rm c}<1$, and since only positive integer $M$ are allowed, we conclude that the
${\rm SU}(4)$ simplex solid states on the cubic lattice are all ordered.  In the mean field theory of Ref.~\onlinecite{arovas:104404},
however, one finds $T_{\rm c}^\ssr{MF}=\zeta/(N^2-1)$, where $\zeta$ is the number of plaquettes associated with a given site.
For the cubic lattice SU(4) model, $\zeta=12$, whence $T_{\rm c}^\ssr{MF}=\frac{4}{5}$, which lies below the actual $T\ns_{\rm c}$.
Thus, the mean field theory {\it underestimates\/} the critical temperature.  In Fig. \ref{su4auto} we show the autocorrelators for the SU(3)
hyperkagome and SU(4) cubic lattice models.  Fig. \ref{su4order} shows the static structure factor and the emergence of Bragg peaks at low temperature.

\begin{figure}[t]
\centering
\includegraphics[width=0.9\linewidth]{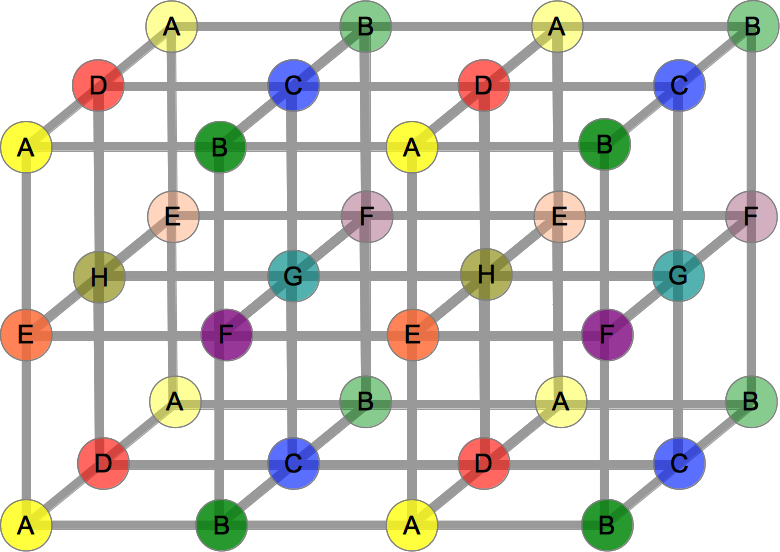}
\caption{A Potts ground state for the SU(8) classical model on the cubic lattice. The magnetic crystal structure is simple cubic.}
\label{su8cubic}
\end{figure}

\begin{figure}[b]
\centering
\includegraphics[width=1.0\linewidth]{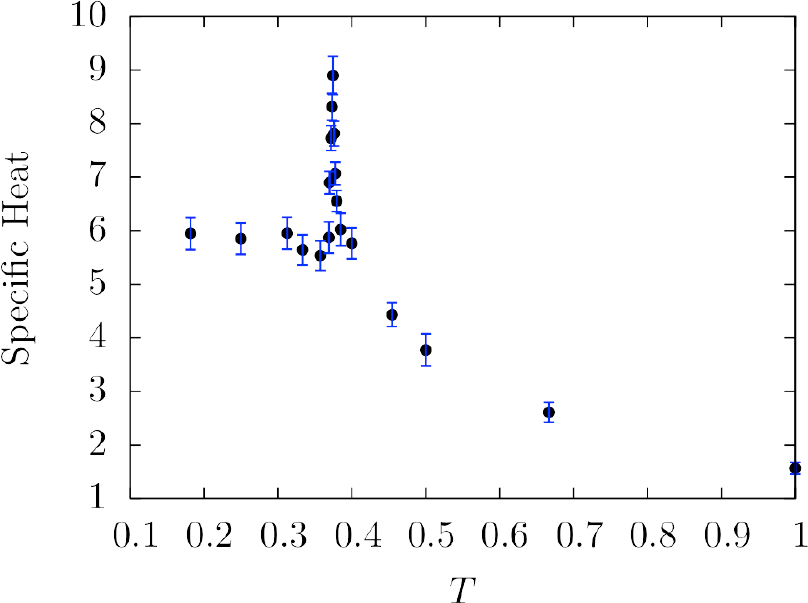}
\caption{Specific heat for the cubic lattice SU(8) model with $N=1000$ sites.  The critical temperature is $T_\Rc\simeq 0.370$.}
\label{su8heat}
\end{figure}

\subsection{SU(8) model on the cubic lattice}
Finally, we consider a three-dimensional model with face-sharing simplices.  On the cubic lattice, with eight species of boson per site,
we can construct the SU(8) singlet operator $\phi\yd_\Gamma$ on each cubic cell.  Each site lies at the confluence of eight such cells,
hence in the state $\sket{\RPsi}=\prod_\Gamma(\phi\yd_\Gamma)^M\sket{0}$, each site is in the fully symmetric representation of SU(8)
described by a Young tableau with one row and $8M$ boxes. Nearest neighbor cubes share a face, next nearest neighbor cubes share
a single edge, and next next nearest neighbor cubes share a single site. The associated classical Hamiltonian for the model is
constructed from eight-site interactions on every elementary cube of the lattice.
\begin{equation}
H_{cl}=-\sum_\Gamma\ln \big|   \eps^{\alpha\nd_1 \cdots \alpha\nd_8}\, z_{\Gamma\nd_1,\alpha\nd_1} \!\!\cdots
z_{\Gamma\nd_8,\alpha\nd_8} \> \big|^2\ ,
\label{HamSU(8)}
\end{equation}
where $\Gamma\nd_i$ are the corners of the elementary cube $\Gamma$. A minimum energy $(E\ns_0=0)$ Potts state can be constructed by
choosing eight mutually orthogonal vectors and arranging them in such a way that corners of every cube are different vectors from this set.
Ground states of this model include all ground states of the eight-state Potts model with eight-spin interactions. Once again, a vast number
of such Potts states is possible. For example, a state with alternating planes, each of them containing only four out of eight Potts spin
directions, has a large number of zero modes. It has a simple cubic pattern, depicted in Fig. \ref{su8cubic}.  We rely on numerical
simulation to determine the preferred state at low temperatures. 

\begin{figure}[t]
\centering
\includegraphics[width=1.0\linewidth]{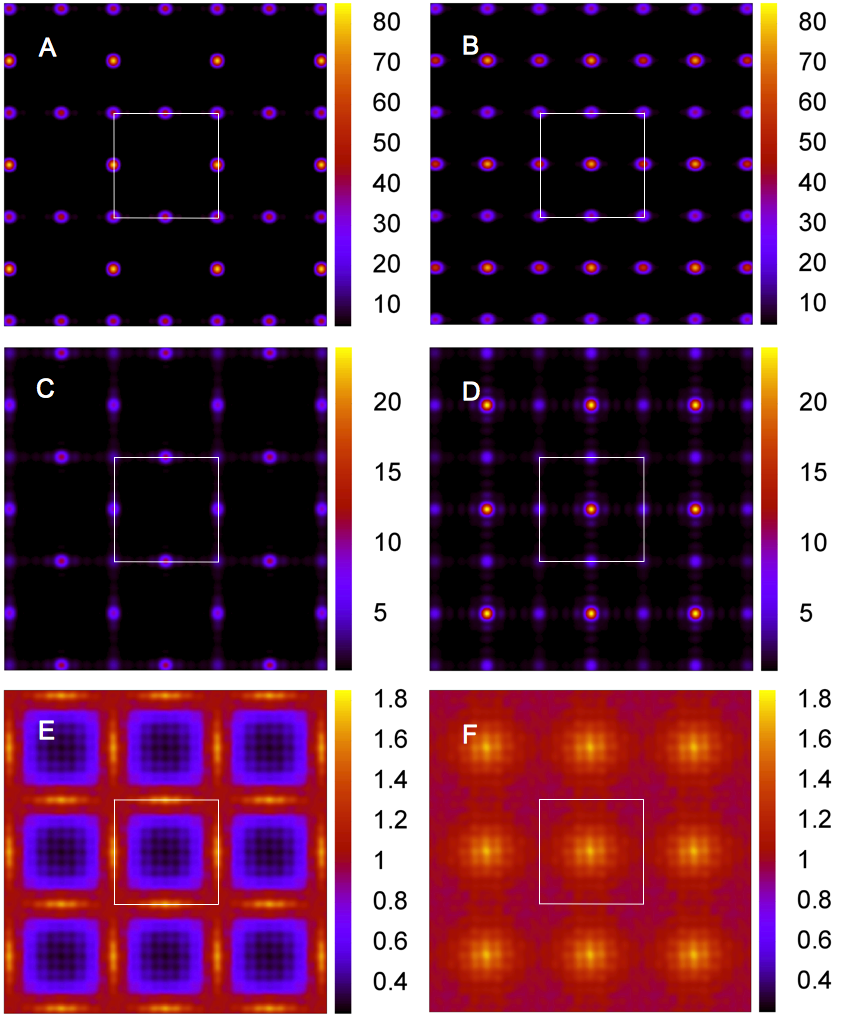}
\caption{Static structure factors for the SU(8) model on the cubic lattice. A and B are $k_z = 0$ and $k_z = \pi / a$ cross sections of the static factor in the ordered phase, $M = 20$.
C and D are $k_z = 0$ and $k_z = \pi / a$ cross sections for higher temperature, M = 2.7, close to the critical value $M_C = 2.67 \pm 0.01$. E and F are $k_z = 0$ and $k_z = \pi / a$
cross sections in the disordered phase, $M = 2$. White lines denote the borders of the Brillouin zone. Note the overall scale; number of sites is $N=1000$.}
\label{su8factors}
\end{figure}

There is a phase transition to the ordered phase at $T\nd_\Rc = 0.370 \pm 0.005$. This is backed by both heat capacity temperature
dependence (Fig. \ref{su8heat}) and static structure factor calculations (Fig. \ref{su8factors}). Our Monte Carlo data for $S(\Bk)$
indicates the presence of spontaneously broken SU(8) symmetry below $T\nd_\Rc$, where Bragg peaks develop corresponding to a simple
cubic structure with a magnetic unit cell which is $2\times 2\times 2$ structural unit cells.
Since $M\ns_\Rc=1/T\ns_\Rc\simeq 2.70$, the SU(8) cubic lattice simplex solid states with $M=1$ and $M=2$ will be quantum disordered,
while those with $M>2$ will have 8-sublattice antiferromagnetic Potts order.  As in the case of the SU(4) model discussed above,
the actual transition temperature is larger than the mean field value $T^\ssr{MF}_\Rc=\zeta/(N^2-1)=\frac{8}{63}=0.127$.

\subsection{The mean field critical temperature}
Conventional wisdom has it that mean field theory always overestimates the true $T\ns_{\rm c}$ because of its neglect of fluctuations.
As discussed in the introduction, in the SU(2) valence bond solid states, the corresponding classical interaction is
$u\ns_{ij}=-\ln\big(\half-\half\nhat\ns_i\cdot\nhat\ns_j\big)$, and one finds $T_{\rm c}^\ssr{MF}=r/3$, where $r$ is the lattice
coordination number. Monte Carlo simulations yield $T\ns_{\rm c}=1.66$ on the cubic lattice ($r=6$, $T_{\rm c}^\ssr{MF}=2$), and
$T\ns_{\rm c}=0.85$ on the diamond lattice ($r=4$, $T_{\rm c}^\ssr{MF}=\frac{4}{3}$) \cite{PhysRevLett.59.799,parameswaran:024408}.
In both cases, the mean field value $T^\ssr{MF}_{\rm c}$ overestimates the true transition temperature.

It is a simple matter, however, to concoct models for which the mean field transition temperature underestimates the actual critical
temperature.  Consider for example an Ising model with interaction $u(\sigma,\sigma')=-\eps^{-1}\ln(1+\eps\sigma\sigma')$, where
the spins take values $\sigma,\sigma'=\pm 1$, and where $0 < \eps < 1$.  If we write $\sigma=\langle\sigma\rangle + \delta\sigma$
at each site and neglect terms quadratic in fluctuations, the resulting mean field Hamiltonian is equivalent to a set of decoupled
spins in an external field $h=rm/(1+\eps m^2)$.  The mean field transition temperature is $T^\ssr{MF}_\Rc=r$, independent of $\eps$.
On the other hand, we may also write $u(\sigma,\sigma')=u\ns_\eps-J\ns_\eps\,\sigma\sigma'$, where $u\ns_\eps=-\ln(1-\eps^2)/2\eps$
and $J\ns_\eps=\eps^{-1}\tanh^{-1}(\eps)$.  On the square lattice, one has $T\ns_\Rc(\eps)=2 J\ns_\eps/\sinh^{-1}(1)$, which diverges
as $\eps\to 1$, while $T^\ssr{MF}_\Rc=4$ remains finite.  For $\eps>0.9265$, one has $T\ns_\Rc(\eps)> T^\ssr{MF}_\Rc$.

Another example, suggested to us by S. Kivelson, is that of hedgehog suppression in the three-dimensional O(3) model.
Motrunich and Vishwanath \cite{Motrunich04} investigated the O(3) model on a decorated cubic lattice with spins present at the vertices
and at the midpoint of each link.  They found $T\ns_\Rc=0.588$ for the pure Heisenberg model and $T^*_\Rc=1.38$ when hedgehogs were
suppressed.  The mean field theory is not sensitive to hedgehog suppression, and one finds $T^\ssr{MF}_\Rc=\frac{2}{\sqrt{3}}=1.15$, which
overestimates $T\ns_\Rc$ but underestimates $T^*_\Rc$.

In both these examples, the mean field partition function includes states which are either forbidden in the actual model, or which
come with a severe energy penalty ($\eps\approx 1$ in our first example).  Consider now the classical interaction derived from the
simplex-solid ground models, $u\ns_\Gamma=-2\ln V\ns_\Gamma$, where
$V\ns_\Gamma=|\eps^{\alpha\ns_1\cdots\alpha\ns_N}z\ns_{\Gamma\ns_1,\alpha\ns_1}\!\!\!\cdots z\ns_{\Gamma\ns_N,\alpha\ns_N}|$
is the internal volume of the simplex $\Gamma$.  If we consider the instantaneous fluctuation of a single spin in the simplex,
we see that there is an infinite energy penalty for it to lie parallel to any of the remaining $(N-1)$ spins, whereas the mean field
Hamiltonian is of the form $H^\ssr{MF}=-\zeta\sum_i h\ns_{\mu\nu}(i)\,Q\ns_{\mu\nu}(i)$, and $h\ns_{\mu\nu}(i)=a\ns_N(m)\delta\ns_{\mu\nu}
+ b\ns_N(m)\,P^{\sigma(i)}_{\mu\nu}$, where $a\ns_N(m)$ and $b\ns_N(m)$ are computed in 
Ref.~\onlinecite{arovas:104404}, and $P^{\sigma(i)}$
is the projector onto the CP$^2$ vector associated with sublattice $\sigma(i)$ in a Potts ground state.  There are no local directions
which are forbidden by $H^\ssr{MF}$, so the mean field Hamiltonian allows certain fluctuations which are forbidden by the true Hamiltonian.
This state of affairs also holds for the SU(2) models, where Monte Carlo simulations found that the mean field transition temperature
overestimates the true transition temperature, as the folk theorem says, but apparently the difference $T\ns_\Rc-T^\ssr{MF}_\Rc$ becomes
positive for larger values of $N$.

\section{Order and Disorder in Simplex Solid States}
To apprehend the reason why the SU(3) hyperkagome model remains disordered for all $T=1/M$ while the SU(4) and SU(8) cubic lattice models
have finite $T$ phase transitions (which in the former case lies in the forbidden regime $T>1$, \ie\ $M<1$), we examine once again
the effective low-temperature Hamiltonian of eqn. \ref{HLT}, derived in Ref.~\onlinecite{arovas:104404},
\begin{equation}
H\nd_{\rm LT}=\sum_\Gamma\sum_{i<j}^N \big|\pi\yd_{\Gamma\nd_i}\,\omega\nd_{\sigma(\Gamma\nd_j)} +
\omega\yd_{\sigma(\Gamma\nd_i)} \pi\nd_{\Gamma\nd_j}\big|^2 \quad.
\end{equation}
The expansion here is about a Potts state, where each simplex $\Gamma$ is fully satisfied such that $V\ns_\Gamma=1$.
In a Potts state, each lattice site $k$ is assigned to a sublattice $\sigma(k)\in\{1,\ldots,N\}$, with $\{\omega\ns_\sigma\}$ a
mutually orthogonal set of $N$ CP$^{N-1}$ vectors and $\pi\yd_i\omega\nd_{\sigma(i)}=0$. It is convenient to take
$\omega\ns_{\sigma,\mu}=\delta\ns_{\mu,\sigma}$, \ie\ the $\mu$ component of the CP$^{N-1}$ vector $\omega\nd_\sigma$ is
$\delta\ns_{\mu,\sigma}$. In $H\nd_{\rm LT}$, the first sum
is over all simplices $\Gamma$, and the second sum is over all pairs of sites $(\Gamma\nd_i,\Gamma\nd_j)$ on the simplex $\Gamma$.

Let us first consider a Potts state which has the same periodicity as the underlying lattice.  In such a state, each simplex corresponds
to a unit cell of the lattice. Examples would include the $\Bq=0$ Potts states of the SU(3) simplex solid on the kagome lattice and the
SU(4) model on the pyrochlore lattice, or a variant of the SU(8) cubic lattice model discussed above, where one sublattice of cubes
is eliminated such that the remaining cubes are all corner-sharing.  In such a structure, we may write $\omega\nd_{\sigma(\Gamma\nd_i)}
\equiv\omega\nd_i$\,, in which case the interaction between sites $i$ and $j$ on the same simplex may be written as
$|\pi^*_{\Gamma\nd_i\,,\,j}+\pi\ns_{\Gamma\nd_j\,,\,i}|^2$, where $\pi\nd_{\Gamma\nd_i\,,\,j}$ is the $j$ component of the $N$-component
vector $\pi\nd_{\Gamma\nd_i}$.  Note that $\pi\nd_{\Gamma\nd_i\,,\,i}=0$.  Since each site is a member of precisely two simplices,
the system may be decomposed into a set of one-dimensional chains, each of which is associated with a pair $(\sigma,\sigma')$ of indices.
Hence there are $\half N(N-1)$ pairs in all.  To visualize this state of affairs, it is helpful to refer to the case of the kagome lattice
in fig. \ref{kagomeQ0}, for which $N=\half N(N-1)=3$.  Thus there are three types of chains: AB, BC, and CA.  Each AB chain is described
by a classical energy function of the form
\begin{align}
H\nd_{{\rm AB}}&=\sum_n \Big( |a^*_n + b\ns_n|^2 + |b^*_n+a\ns_{n+1}|^2\Big)\\
&=\sum_k\begin{pmatrix} a^*_k & b\ns_{-k} \end{pmatrix} \begin{pmatrix} 2 & 1+ e^{-ik} \\ 1+e^{ik} & 2 \end{pmatrix}
\begin{pmatrix} a\ns_k \\ b^*_{-k} \end{pmatrix}\quad.\nonumber
\end{align}
This yields two excitation branches, with dispersions $\omega\ns_\pm(k)=2\pm 2\cos(\half k)$.
Thus we recover $N(N-1)$ complex degrees of freedom, or $2N(N-1)$ real degrees of freedom, per unit cell, as derived in \S \ref{cdof}.

In Ref.~\onlinecite{arovas:104404}, the fixed length constraint of each CP$^{N-1}$ vector $z\nd_i$ was approximated by implementing
the nonholonomic constraint $\langle\pi\yd_i\pi\nd_i\rangle \le 1$, which in turn is expressed as
$|\chi|^2+\langle\pi\yd_i\pi\nd_i\rangle=1$, where $\chi$ plays the role of a condensate amplitude. This holonomic constraint is
enforced with a Lagrange multiplier $\lambda$, so that the free energy per site takes the form of eqn. \ref{ferg}, where $g(\ve)$
is the total density of states per site, normalized such that $\int\limits_0^\infty\! d\ve\>g(\ve)=1$. For the models currently under
discussion, we have $g(\ve)=g\nd_\ssr{1D}(\ve)$, where
\begin{equation}
\begin{split}
g\nd_\ssr{1D}(\ve)&=\int\limits_0^{2\pi}\!{d\theta\over 2\pi}\>\delta(2-2\cos\theta-\ve)\\
&={\Theta\big(2-|\ve-2|\big)\over \pi\sqrt{\ve(4-\ve)}}\quad,
\end{split}
\end{equation}
characteristic of one-dimensional hopping.  The spectrum is confined to the interval $\ve\in[0,4]$, and extremizing
with respect to $\lambda$ yields the equation
\begin{equation}
1=|\chi|^2+(N-1)\,T\!\int\limits_0^\infty\!\!d\ve\>{g(\ve)\over\ve+\lambda}\quad.
\end{equation}
If $\int\limits_0^\infty\!d\ve\>\ve^{-1} g(\ve) <\infty$, then $\lambda=0$ and $|\chi|^2>0$. This is the broken SU($N$) symmetry regime.
Else, $\lambda>0$ and $\chi=0$, corresponding to a gapped, quantum disordered state.

\subsection{SU(3) kagome and hyperkagome models}
For the SU(3) kagome and hyperkagome models, expanding about a $\Bq=0$ Potts state, the free energy per site for the low temperature
model $H\ns_{\rm LT}$, implementing the nonholonomic mean fixed length constraint for the CP$^2$ spins, is found to be
\begin{equation}
f(T,\lambda)=-\lambda+2T\ln\!\Bigg({2+\lambda+\sqrt{\lambda(\lambda+4)}\over 2T}\Bigg)\quad.
\end{equation}
Setting $\partial f/\partial\lambda=0$ yields $\lambda=2\big(\sqrt{1+T^2}-1\big)$. These systems are in gapped, disordered phases for
all $T$, meaning that the corresponding quantum wave functions are quantum-disordered for all values of the discrete parameter $M$.
The low temperature specific heat is $C(T)=2-2T+{\cal O}(T^2)$.

\begin{figure}
\includegraphics[width=0.47\textwidth]{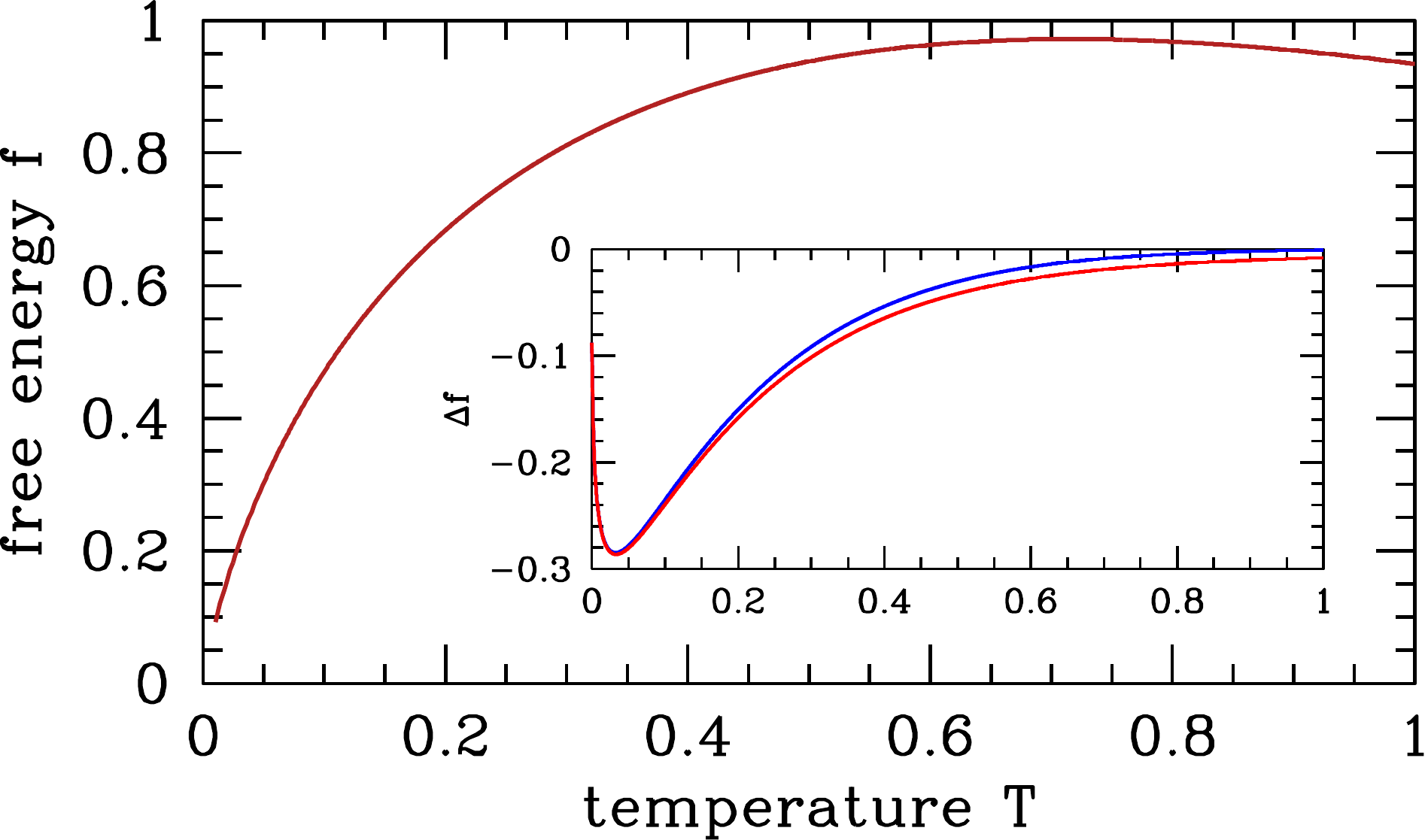}
\caption{Free energy per site for the $\Bq=0$ states of the SU(3) kagome and hyperkagome lattice models.  The inset shows the difference
in free energies $\Delta f$ between the $\sqrt{3}\times\sqrt{3}$ structure on the kagome lattice and the $\Bq=0$ state (blue), and
the corresponding difference for the analogous state in the hyperkagome lattice (36 site magnetic unit cell).}
\label{fig:KHKferg}
\end{figure}

In the $\sqrt{3}\times\sqrt{3}$ state on the kagome lattice, we have\cite{arovas:104404}
\begin{equation}
g\nd_{\rm K}(\ve)=\frac{1}{6}\,\Big\{\delta(\ve)+2\,\delta(\ve-1) + 2\,\delta(\ve-3) + \delta(\ve-4)\Big\}\ ,
\end{equation}
whereas for the analogous structure in the hyperkagome lattice, with a 36 site magnetic unit cell, we find
\begin{align}
g\nd_{\rm HK}(\ve)&=\frac{1}{12}\,\Big\{2\,\delta(\ve) + \delta(\ve-1) + 2\,\delta(\ve-2) + \delta(\ve-3)\nonumber\\
&\qquad\quad+2\,\delta(\ve-4)+\delta(\ve-1-\phi)+\delta(\ve-2+\phi)\nonumber\\
&\qquad\quad+\delta(\ve-2-\phi)+\delta(\ve-3+\phi)\Big\}\quad,
\end{align}
where $\phi=\half\big(1+\sqrt{5}\big)\simeq1.618$.  For the kagome system, we obtain
\begin{equation}
{1\over T}={2u\over 3}\cdot\bigg\{{1\over u^2-4}+{2\over u^2-1}\bigg\}\quad,
\end{equation}
where $u\equiv\lambda+2$.  For the hyperkagome system, 
\begin{align}
{1\over T}&={u\over 3}\cdot\bigg\{ {2\over u^2-4} + {1\over u^2-\phi^2} \\
&\qquad\quad+{1\over u^2-1} + {1\over u^2-(1-\phi)^2} + {1\over u^2}\bigg\}\nonumber\ .
\end{align}
One then obtains $\lambda\ns_{\rm K}=\frac{1}{3} T + \frac{35}{108} T^2+{\cal O}(T^3)$ for kagome
$\lambda\ns_{\rm HK}=\frac{1}{3} T + \frac{31}{108} T^2+{\cal O}(T^3)$ for hyperkagome, at low temperatures.
The corresponding specific heat functions are then
\begin{equation}
\begin{split}
C\nd_{\rm K}(T)&=\frac{5}{3} - \frac{35}{54}\, T + {\cal O}(T^2)\\
C\nd_{\rm HK}(T)&=\frac{5}{3} - \frac{31}{54}\, T + {\cal O}(T^2)\ .
\end{split}
\end{equation}
Both tend to the same value as $T\to 0$.  For the kagome system, we found $C(0)=1.84\pm 0.03$, close to the value of $\frac{16}{9}$
obtained by augmenting the quadratic mode contribution of $\frac{5}{3}$ with that from the quartic modes, whose contribution is
$\Delta C=\frac{1}{9}$.  Our hyperkagome simulations, however, found $C(0)\approx 2$, with no apparent deficit from zero modes
or quartic modes.  Again, this is consistent with the structure factor results, which show no hint of any discernible structure
down to the lowest temperatures.  A plot of the free energy per site for the $\Bq=0$ Potts state on the kagome and hyperkagome lattices,
and the free energy difference per site between this structure and the $\sqrt{3}\times\sqrt{3}$ kagome structure and its hyperkagome analog
are shown in Fig. \ref{fig:KHKferg}.

\subsection{SU(4) cubic lattice model}
We now analyze the low-energy effective theory of the SU(4) cubic lattice model, expanding about the Potts state depicted in 
fig. \ref{bcc_cubic}.  The magnetic unit cell consists of four sites.  Let the structural cubic lattice constant be $a\equiv 1$.
The magnetic Bravais lattice is then BCC, with elementary direct lattice vectors
\begin{equation*}
\Ba\ns_1=(1,1,1)\quad,\quad\Ba\ns_2=(-1,1,1)\quad,\quad\Ba\ns_3=(1,-1,1)
\end{equation*}
and elementary reciprocal lattice vectors
\begin{equation*}
\Bb\ns_1=(\pi,\pi,0)\quad,\quad\Bb\ns_2=(-\pi,0,\pi)\quad,\quad\Bb\ns_3=(0,-\pi,\pi)\ .
\end{equation*}
In the Potts state, the A sites lie at BCC Bravais lattice sites $\BR$, with B sites at $\BR+\xhat$, C at $\BR+\yhat$, and D at
$\BR+\zhat$.  There are $2(N-1)=6$ real degrees of freedom per lattice site, and hence 24 per magnetic unit cell.
The low temperature Hamiltonian may be written as a sum of six terms
\begin{equation}
H\ns_{\rm LT}=H\ns_\ssr{AB}+H\ns_\ssr{AC}+H\ns_\ssr{AD}+H\ns_\ssr{BC}+H\ns_\ssr{BD}+H\ns_\ssr{CD}\quad,
\label{HABCD}
\end{equation}
where $H\ns_\ssr{AB}$ couples the B component of the $\pi$ vector on the A sites with the A component of the $\pi$ vector on the B sites.
Explicitly, we note that an A site at $\BR$ has B neighbors in unit cells at $\BR$, at $\BR-\Ba\ns_1$, at $\BR+\Ba\ns_2$, at $\BR-\Ba\ns_3$,
at $\BR-\Ba\ns_1+\Ba\ns_2$, and at $\BR-\Ba\ns_1+\Ba\ns_2+\Ba\ns_3$.  Thus,
\begin{align}
H\ns_\ssr{AB}&=\sum_\BR\Big\{ b^*_\BR\big(a\ns_\BR+a\ns_{\BR-\Ba\ns_1}+a\ns_{\BR+\Ba\ns_2} + a\ns_{\BR-\Ba\ns_3}
+a\ns_{\BR-\Ba\ns_1+\Ba\ns_2}\nonumber\\
&\quad+a\ns_{\BR-\Ba\ns_1+\Ba\ns_2+\Ba\ns_3}\big)+\hbox{\rm c.c.}+6|a\ns_\BR|^2 + 6|b\ns_\BR|^2\Big\}\nonumber\\
&=6\sum_\Bk \begin{pmatrix} a^*_\Bk & b\ns_{-\Bk} \end{pmatrix} \begin{pmatrix} 1 & \gamma\ns_\Bk \\ \gamma^*_\Bk & 1 \end{pmatrix}
\begin{pmatrix} a\ns_\Bk \\ b^*_{-\Bk} \end{pmatrix}\quad ,
\end{align}
where
\begin{align}
\gamma\ns_\Bk&=\frac{1}{3} \, e^{i(\theta\ns_2-\theta\ns_1)/2}\Bigg\{ \cos\!\bigg({\theta\ns_1-\theta\ns_2\over 2}\bigg)
+\cos\!\bigg({\theta\ns_1+\theta\ns_2\over 2}\bigg)\nonumber\\
&\hskip1.2in+\cos\!\bigg({\theta\ns_1-\theta\ns_2\over 2}-\theta\ns_3\bigg)\Bigg\}
\end{align}
with $\Bk=\frac{1}{2\pi}\sum_{i=1}^3\theta\ns_i\,\Bb\ns_i$. This leads to two bands, with dispersions
$\omega\ns_\pm(\Bk)=6\big(1\pm|\gamma\ns_\Bk|\big)$.  All the other Hamiltonians on the RHS of eqn. \ref{HABCD} yield the same dispersion.
Counting degrees of freedom, we have four real (two complex) modes
per $\Bk$ value (${\rm Re}\,a\ns_\Bk$, ${\rm Im}\,a\ns_\Bk$, ${\rm Im}\,b\ns_\Bk$ and ${\rm Im}\,b\ns_\Bk$), and six independent Hamiltonians
on in eqn. \ref{HABCD}, corresponding to 24 real modes per unit cell, as we found earlier.  The bottom of the $\omega\ns_-(\Bk)$
band lies at $|\gamma\ns_\Bk|=1$, which entails $\theta\ns_1=\theta\ns_2=\theta\ns_3=0$.  Expanding about this point, the dispersion
is quadratic in deviations, corresponding to the familiar bottom of a parabolic band.  The density of states is then
$g(\ve)\propto\sqrt{\ve}$, which means that $\lambda=0$ and $|\chi(T)|^2$ interpolates between $|\chi(0)|^2=1$ and
$|\chi(T\ns_\Rc)|^2=0$, where
\begin{equation}
T\ns_\Rc={1\over (N-1)\int\limits_0^\infty\!d\ve\>\ve^{-1}g(\ve)}
\end{equation}
is the prediction of the low energy effective theory. Because the low-temperature effective hopping theory for edge-sharing
(and face-sharing) simplex solids involves fully three-dimensional hopping, the band structure of their low-lying excitations
features parabolic minima, which in turn permits a solution with $\lambda\ne0$, meaning the ordered state is stable over a range
of low temperatures.  We find $T\ns_\Rc=1.978$ for the edge-sharing simplex solid model on the simple cubic lattice.
This is substantially greater than both the mean field result $T_\Rc^\ssr{MF}=\frac{4}{5}$ and the Monte Carlo
result $T\ns_\Rc\simeq 1.485$.

\begin{figure}
\includegraphics[width=\columnwidth]{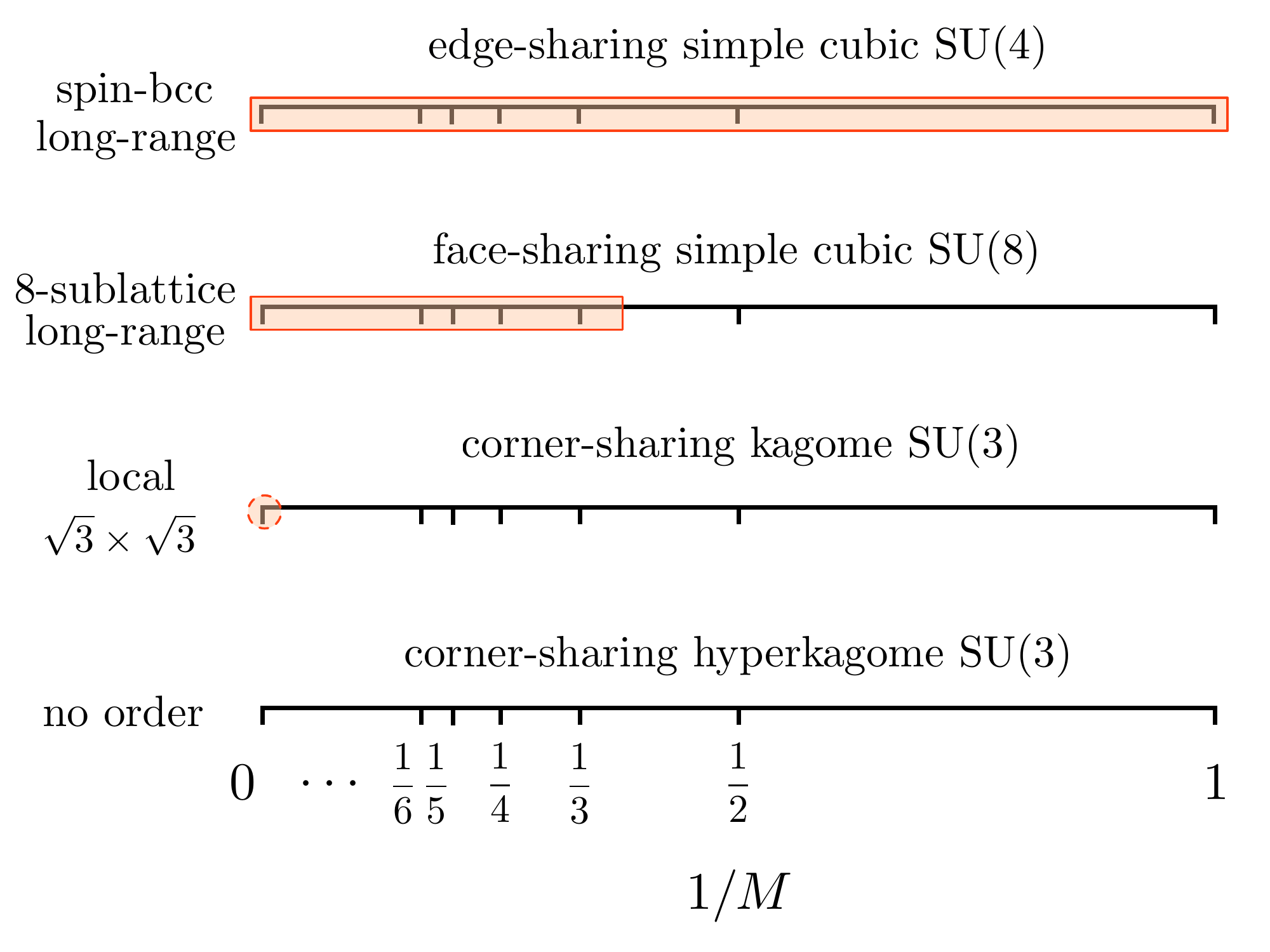}
\caption{\label{fig:fig_PD} {\bf Structure of simplex solids as a function of discrete parameter $M$.} The parameter range for which long-range (local) order emerges is shaded and bounded by solid (dashed) lines; a brief description of the order is also given.  Whereas on the cubic lattice  the edge-sharing SU($4$) model is always long-ranged ordered, the face-sharing SU($8$) model has quantum-disordered ground states for $M=1,2$. The SU($3$) model exhibits quantum disorder for all $M$, with local $\sqrt{3}\times\sqrt{3}$ correlations strengthening as $M\rightarrow \infty$ on the kagome lattice while on the hyperkagome, no local or long range order is apparent at any $M$.}
\end{figure}

\section{Concluding Remarks}

We have studied the structure of exact simplex solid ground states of SU($N$) spin models, in two and three dimensions,
via their corresponding classical companion models that encode their equal time correlations. The discrete parameter
$M$ which determines the on-site representation of SU($N$) sets the temperature $T=1/M$ of each classical model, which
then may be studied using standard tools of classical statistical mechanics.  Our primary
tool is Monte Carlo simulation, augmented by results from mean field and low-temperature effective theories.  This work
represents an extension of earlier work on SU(2) AKLT models.

Through a study of representative models with site-, edge-, and face-sharing simplices, we identify three broad categories
of simplex solids, based on the $T$-dependence of the associated classical model: 

\begin{enumerate}

\item[1.] Models which exhibit a phase transition in which SU($N$) is broken at low temperature, corresponding
to a classical limit $M\to\infty$ analogous to $S\to\infty$ for SU(2) systems, as exemplified by the edge-sharing
SU(4) and face-sharing SU(8) cubic lattice simplex solids. Whether or not these models have quantum-disordered for physical (i.e., integer) values of the singlet parameter $M$ depends on the precise value of the transition temperature.

\item[2.] Models which exhibit no phase transition down to $T=0$, but reflect strong local ordering which breaks
lattice and SU($N$) symmetries, as in the SU(3) model on the kagome lattice.  While the low and high $M$ limits of
these simplex solids appear to be in the same (quantum-disordered) phase, we expect the ground state expectation values for $M\to\infty$
are dominated by classical configurations with a large density of local zero modes.

\item[3.] Models which exhibit neither a phase transition nor apparent local order down to $T=0$ and are hence quantum-disordered and featureless for all $M$. These simplex solids
perhaps best realize the original AKLT ideal of a featureless quantum-disordered paramagnet, for the case of SU($N$) spins.  The hyperkagome lattice SU(3)
simplex solid is representative of this class.

\end{enumerate}  
These results are summarized graphically in Fig.~\ref{fig:fig_PD}.

The parent Hamiltonians which admit exact simplex solid ground states are baroque and bear little
resemblance to the simple SU($N$) Heisenberg limit typically studied.  Nevertheless, we may regard the simplex solids
as describing a phase of matter which may include physically relevant models.  This state of affairs obtains in $d=1$,
where the AKLT state captures the essential physics of the $S=1$ Heisenberg antiferromagnet in the Haldane phase.
We also note that SU$(N)$ magnetism, once primarily a theorists' toy, may be relevant in certain experimental
settings; in this context, there has been recent progress examining the feasibility of realizing such generalized spin models with systems of ultracold atoms, particularly those involving alkaline earth atoms ~\cite{PhysRevLett.91.186402, Gorshkov:2010fk}. Whether the states analyzed in this paper will find a place in the phase diagrams of such systems remains an open question, that we defer to the future.

\section*{Acknowledgments} We are grateful to S. Kivelson and J. McGreevy for very helpful discussions and suggestions, and to S.L. Sondhi for collaboration on prior related work (Ref.~\onlinecite{parameswaran:024408}). SAP acknowledges several illuminating discussions with I. Kimchi on examining featureless phases via plasma mappings, and the hospitality of UC San Diego and the Institute of Mathematical Sciences, Chennai, where parts of this work were completed. This work was supported in part by NSF grant DMR 1007028 (YYK, DPA), by UC Irvine start-up funds and the Simons Foundation (SAP). 
\bibliography{hkagrefs}
\end{document}